\documentclass[journal]{IEEEtran}
\ifCLASSINFOpdf
\else
\fi
\usepackage[all=normal,paragraphs=tight,wordspacing=normal,indent=normal,floats=tight,leading=normal,lists=tight]{savetrees}
\usepackage {graphicx}
\usepackage{gensymb}
\usepackage{multirow}
\usepackage[usenames,dvipsnames,svgnames,table]{xcolor}
\usepackage{algorithm,algpseudocode}
\usepackage {mathptm}
\usepackage {amssymb,amsmath}
\usepackage{algorithm,algpseudocode}
\usepackage {graphicx}
\usepackage {url}
\usepackage{xcolor}
\usepackage{multirow}
\usepackage{epstopdf}
\usepackage{mathtools}
\usepackage{amsmath}
\usepackage{gensymb}
\usepackage[english]{babel}
\usepackage{microtype}

\begin{document}
\title{Test and  Yield Loss Reduction of AI and Deep Learning Accelerators}

\author{Mehdi~Sadi,~\IEEEmembership{Member,~IEEE,}
        and~Ujjwal~Guin,~\IEEEmembership{Member~,~IEEE}
\thanks{This work was supported in part by the IGP grant from Auburn University, Auburn, AL.}
\thanks{Mehdi Sadi and Ujjwal Guin are with the Department of Electrical and Computer Engineering, Auburn University, Auburn, AL 36849, USA (e-mail: mehdi.sadi@auburn.edu, ujjwal.guin@auburn.edu)}
\thanks{Manuscript received August 05, 2020; revised October 11, 2020; accepted XX XX, 20XX.}}

\markboth{Journal of \LaTeX\ Class Files,~Vol.~XX, No.~X, August~202X}%
{Sadi \MakeLowercase{\textit{et al.}}: Test and Yield Loss Reduction of AI and Deep Learning Accelerators}
\maketitle

\begin{abstract}
With data-driven analytics becoming mainstream, the global demand for dedicated AI and Deep Learning accelerator chips is soaring. These accelerators, designed with densely packed Processing Elements (PE), are especially vulnerable to the manufacturing defects and functional faults common in the advanced semiconductor process nodes resulting in significant yield loss. In this work, we demonstrate an application-driven methodology of binning the AI accelerator chips, and yield loss reduction by correlating the circuit faults in the PEs of the accelerator with the desired accuracy of the target AI workload. We exploit the inherent fault tolerance features of trained deep learning models and a strategy of selective deactivation of faulty PEs to develop the presented yield loss reduction and test methodology. An analytical relationship is derived between fault location, fault rate, and the AI task's accuracy for deciding if the accelerator chip can pass the final yield test. A yield-loss reduction aware fault isolation, ATPG, and test flow are presented for the multiply and accumulate units of the PEs. Results obtained with widely used AI/deep learning benchmarks demonstrate that the accelerators can sustain 5\% fault-rate in PE arrays while suffering from less than 1\% accuracy loss, thus enabling product-binning and yield loss reduction of these chips.
\end{abstract}

\IEEEpeerreviewmaketitle

\section{Introduction}
\label{sec:Intro}

The demand for Artificial Intelligence (AI) and deep learning is growing at a rapid pace across a wide range of applications such, as self-driving vehicles, image and voice recognition, medical imaging and diagnosis, finance and banking, natural resource explorations, defense operations, etc. Because of these data-driven analytics and AI boom, demands in deep learning and AI will emerge at both data centers and the edge \cite{AA1}-\cite{M1}.  In a recent market research \cite{M1}, it has been reported that AI-related semiconductors will see a growth of about 18 percent annually over the next few years - five times greater than the rate for non-AI applications. By 2025, AI-related semiconductors could account for almost 20 percent of all semiconductor demand, which would translate into about \$67 billion in revenue \cite{M1}.

Although GPU was adopted by the AI community, by design GPUs were not optimized for AI workloads \cite{M1}\cite{gpu}. As a result significant R\&D efforts in developing AI accelerators - optimized to achieve much higher throughput in deep learning compared to GPUs  -  are underway from academia \cite{AA2}\cite{AA3}, big techs \cite{AA4}-\cite{AA6}, as well as startups \cite{AA7}\cite{cb}. Dedicated accelerators are in high demand for both the cloud-based training, and inference tasks on edge devices. The training procedure is time-consuming as it requires many learning samples to adapt the network parameters. For instance, a self-driving car’s neural network has to be trained with many images of possible objects it can encounter on the road, and this will require multiple high-performance AI accelerators on the cloud. During inference, AI algorithms handle less data but rapid responses are required as they are often used in critical time-sensitive applications. For example, an autonomous vehicle has to make immediate decisions on objects it sees during driving, a medical device must interpret a trauma patient’s brain scans immediately. As a result, high throughput accelerators running on edge devices and capable of fast inference are required. In AI technology innovation and leadership, high-throughput AI accelerator hardware chips will serve as the differentiator \cite{M1}\cite{cb}.

Millions of Multiply and Accumulate (MAC) operations are needed in modern AI tasks, for example, AlexNet \cite{Time} and ResNET-50 \cite{res} require 0.7 Billion and 3.7 Billion MAC operations, respectively, to classify a single image from the ImageNet \cite{img} dataset. Modern AI accelerators contain thousands of Processing Elements (PE) distributed in densely packed arrays in a single chip/die \cite{AA1}-\cite{AA5} or in a multi-chiplet based chip \cite{cb}. To accommodate as many PEs as possible in the AI accelerator, a large chip (e.g., wafer-scale) would be the best solution \cite{cb}.  However, manufacturing large chips is difficult due to the long interconnect wires and the possibility of particle defects. Aggressive scaling of design rules \cite{Y2}, lithography imperfections \cite{Y4}, local edge roughness, interconnect pitch reduction, etc., in 10nm and newer semiconductor technologies \cite{Y2} have caused yield (i.e., the fraction of total manufactured chips that can be sold to the customer) to become as important as the conventional design metrics of Power, Performance, Area (PPA) for the process to be economically viable \cite{Y1}\cite{Y3}. Moreover, internal cell defects have become a major yield limiter because of aggressive scaling of design rules and lithography limitations in printing the features, giving rise to cell-aware test \cite{CA1}. In addition to the regular stuck-at and timing faults at the cell I/O level, at advanced technologies, the transistor-level and other cell-internal faults need to be added to the fault list as yield has become more vulnerable to internal cell defects \cite{CA2}. Although yield data of semiconductor process are considered a well-guarded trade secret and not published, it is well known that yield loss has caused significant delays in product readiness, and loss in market share and revenue for a leading processor manufacturer at 10nm and 7nm \cite{int}. To integrate more PE in the accelerator, a two-level chiplet \cite{ chiplet1} based approach, where many PE are placed on a smaller chiplet, and then multiple chiplets are connected with silicon interconnect fabric \cite{ chiplet2} to form the accelerator is a viable solution. Although a relatively smaller size would minimize particle-induced random defects on upper metal layers in the chiplets and improve defect-induced yield loss, the individual PEs - internal to the chiplet - will still be susceptible to systematic defects \cite{Y1} and lithography imperfections \cite{Y2} that impact the transistor layers or the ultra-dense lower metal layers.

As many PEs are densely placed in the AI accelerators, defects and circuit faults are likely to occur in some PEs. Fortunately, the stochasticity inherent in the backpropagation-based training of Deep Neural Networks (NN) and Deep Convolutional Neural Networks (CNN) - the primary building blocks of AI systems - offers a certain degree of resilience and error tolerance to deep learning tasks. Moreover, the intelligent application of techniques such as dropout, pruning, and quantization during training can further increase the robustness of a trained NN/CNN against variations and noise during inference \cite{P1}-\cite{D1}. The error-resilience properties of well-trained deep NN/CNNs can be exploited in hardware by allowing the PE of the dense AI accelerator to incur some circuit faults - caused by semiconductor manufacturing process variation induced defects - and still function correctly within an accuracy bound. As a result, a fault-tolerance aware test flow is required for these accelerators that can test the individual PEs, and certify if the fault of the PE is acceptable or unacceptable depending on how many of the rest of the PEs are fault-free or faulty, and the impact of faults on AI accuracy. An innovative solution would be to implement a fine-grained fault tolerance scheme that allows the deactivation of individual PEs in the event that the PE fault rate (i.e., the fraction of PEs that are faulty) exceeds a threshold. Following this approach, an AI accelerator with some faulty PEs to still function, and will not cause the discard of the whole AI accelerator chip, resulting in a significant reduction in yield loss \cite{Y2}-\cite{Y3}.

In this paper, we propose \textit{\textbf{YAOTA}: \textbf{Y}ield and \textbf{A}ccuracy aware \textbf{O}ptimum \textbf{T}est of \textbf{A}I accelerators}, which considers the accuracy-sensitivity and fault-tolerance of AI applications into test pattern generation for the accelerators in deciding whether it will pass the yield test. The key contributions and highlights of this paper are as follows,
\begin{itemize}
\item    An analytical relationship is established - based on the actual AI workload to be executed - between the (i) faults of the MAC modules, (ii) the rate of faults, and the accuracy of the AI task. This relationship is used in demonstrating that AI accelerator chips can still function correctly despite having few faulty PEs, thus enabling product-binning and yield saving.
\item    An accuracy-aware fault isolation and test pattern generation methodology is presented to group the MAC faults by their logic cones into categories: (i) critical (unacceptable), and (ii) non-critical (acceptable), according to their impact on the accuracy of AI workload. Responses from these test patterns dictate yield decisions - whether to ship to the customer at reduced throughput, or discard as yield loss. Using the reduced test pattern set for critical faults, only the critical faults of the AI accelerator can be tested as a quick method to assess the yield. Next, the chips that passed the first yield-test can be tested for non-critical faults to grade those into different speed/throughput bins. The accelerators in the top bin (i.e., without any fault) can be sold at a premium price for safety-critical applications such as self-driving cars, whereas accelerators in the lower bins (with few faults, e.g., less than 5\% fault rate) can be used in other AI/deep-learning tasks that can tolerate errors with minimal performance loss. 
\item    A strategy of fault-aware training and selective deactivation of faulty PEs during inference is presented to minimize the accuracy loss due to faulty MACs.
\item Simulation results from 50,000 image samples on widely used CNN (AlexNet, ResNet-50, VGG16, LeNet5) \cite{CNN}, and 10,000 data samples on different NN architectures are presented to demonstrate the relationship between fault rate and accuracy. Results show that with 5\% fault rate the normalized accuracy of NN and CNN only degrade by less than 1\%.
\end{itemize}

The rest of the paper is organized as follows. Related work and AI/Deep learning accelerator backgrounds are covered in Section II.  The proposed YAOTA methodology, test flow, and hardware control scheme are presented in Section III. Simulation results are presented in Section IV followed by conclusions in Section V.

\section{Background}
\subsection{Related Work}
\vspace{-0.05in}	
\label{sec:related}
The error-tolerance nature of AI algorithms has been exploited in hardware to design energy-efficient AI accelerator architectures with approximate MACs \cite{A1}-\cite{A4}. In  \cite{A1}\cite{A2}, using benchmark-driven analysis each neuron was ranked according to its sensitivity and error contribution to the output, and neurons that contributed the least to the error were then approximated and the network was retrained to recover accuracy loss. In hardware, neurons that were approximated were assigned to approximate PEs, while others were assigned to exact PEs \cite{A1}\cite{A2}. However, the challenges with these approximate approaches are, (i) sensitivity-based sorting, approximation, and retraining of neurons are always dependent on the workload \cite{A2}\cite{A4}. (ii) Assigning less sensitive neurons to approximate PEs will require runtime hardware reconfiguration for different tasks. (iii) In \cite{A1}-\cite{A4} pruning was not considered, but in modern NN/CNN pruning is applied as a well-established method to reduce network size where less important/sensitive connections are already pruned/removed during training \cite{P1}-\cite{P3}. Hence, the techniques of \cite{A1}-\cite{A4} will be much less effective when applied on an already pruned NN/CNN. (iv) Most importantly, since these approximate CNN/NN assume that the only source of error is the deterministic approximate multiplier and adders, any additional faults in the PE/MAC from process variation induced defects will cause a significant amount of inaccuracy in the prediction during inference.

With the widespread use application AI accelerators, the testing of these hardware has become an emerging research problem \cite{T3}\cite{NY}.  In \cite{T3}, a comprehensive structural test flow was proposed that first identified critical faults by comparing the accuracies of the exact fault-free gate-level circuit of the neural network, and that of a faulty version. Next, the entire circuit was converted into an Boolean satisfiability (SAT) instance and solved with SAT solver with test patterns for the critical faults only. Since this approach is expensive and not scalable, a functional test method was proposed \cite{T3}, where real workloads - the test images from MNIST and CIFAR10 benchmarks - are applied as test input to the gate-level netlist of the implemented neural network. To find if a fault was critical or ignorable, the fault was first injected into a neuron module and the test image set was applied. If the prediction accuracy of the test image set was within a certain threshold, the fault was considered unimportant, otherwise, it was a critical fault.  However, the proposed simple approach of creating RTL of the full neural network, followed by gate-level synthesis and fault injection may not be scalable for mainstream CNNs \cite{CNN} such as AlexNet, VGG, ResNet, MobileNet, etc. that are used for real-world computer vision tasks of image recognition with many classes \cite{AA1}\cite{AA2}\cite{img}. 

In \cite{NY} the authors first identified the location of faulty MACs (i.e., $MAC_{i,j}$),  in the TPU systolic array and then all weights that were mapped to those faulty MACs were pruned (i.e., $\forall w_{i,j} = 0$,where $i,j$=location of faulty MAC), followed by retraining. Extra bypass MUX was required for each MAC to bypass it in case it was faulty. However, the major challenges of this approach are, (i) the pruning with retraining technique can preserve the model accuracy when only the small magnitude weights are pruned \cite{P1}-\cite{P3}. If a large magnitude weight is mapped to a faulty MAC, this method will result in accuracy degradation. (ii) Moreover, in modern AI/deep learning the models are already pruned \cite{P1}-\cite{P3} to reduce energy consumption and storage requirements, as a result further pruning of weights  - corresponding to the faulty MACs - of an already pruned model will significantly reduce its accuracy. (iii) Another drawback of \cite{NY}  is that weight pruning and retraining needs to be performed for each TPU chip based on its unique fault map and each individual workload. 

\textcolor{black}{ In contrast to permanent stuck-at  faults, for low-energy operations where voltage is scaled down, possible timing \cite{TE} and memory errors \cite{ME} can also occur. However, by using appropriate PV guard-band, using circuits with shorter critical-paths and memory ECC these faults can be avoided.} 

\subsection{AI/Deep Learning Accelerator Architecture and Processing Element Faults}
\label{sec:Arch}
\vspace{-0.05in}
\begin{figure}[h]
\centering
\includegraphics[width=2.8in,height=2.6in]{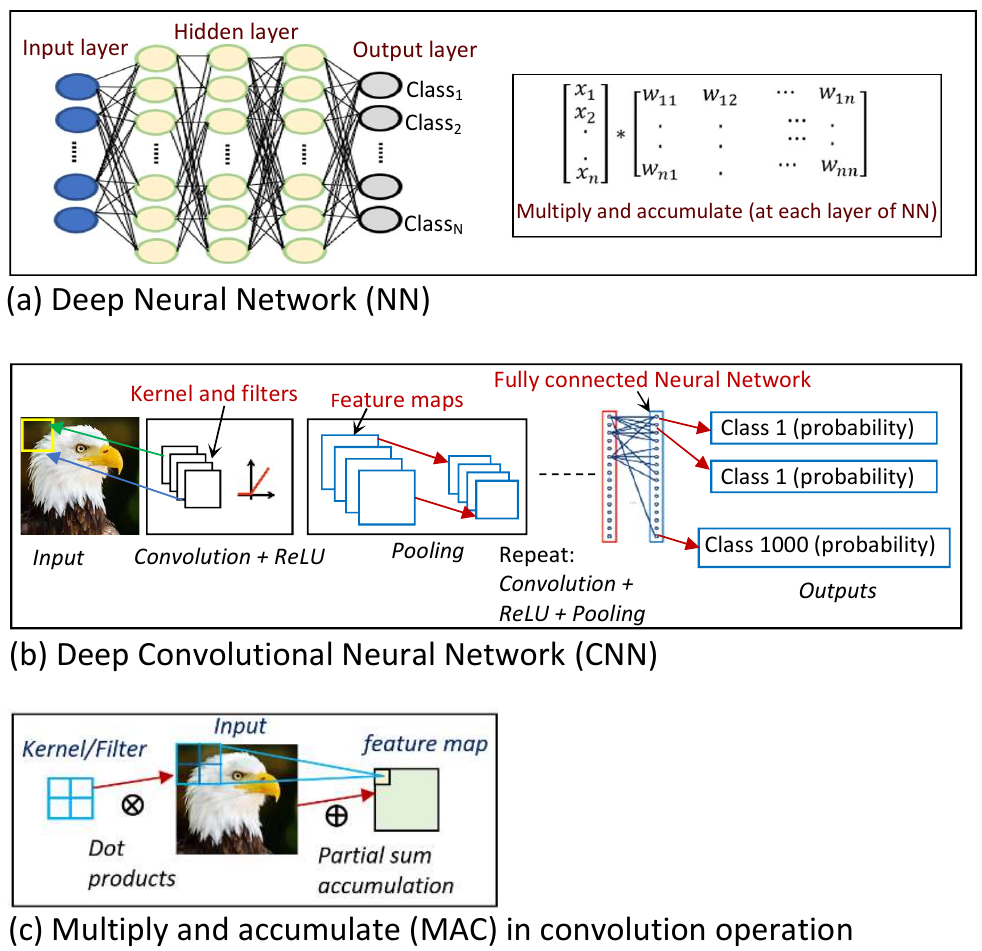}
\caption {(a) Deep NN; (b) Deep CNN; (c) Convolution in CNN \cite{AA1}\cite{CNN}.}
\label{fig:cn}
\vspace{-0.05in}
\end{figure}

At the essence of AI/Deep Learning algorithms are the backpropagation-based NN and CNN. As shown in Fig. \ref{fig:cn} (a), a deep NN consists of an input layer, followed by several hidden layers and a final output layer. Depending on the data size, the complexity of training, dropout, and pruning rate, some layers in the NN are fully connected and others sparsely connected \cite{P1}-\cite{D1}. The connection strengths between the adjacent layers are represented by a weight matrix $W$, and the matrix parameters $w_i$  are learned by the backpropagation equation, $\Delta w_i=-a*\frac{\partial Error}{\partial w_i}$, where $a$ is learning rate and $Error$ is the prediction error. During the forward pass of training and inference phases, the output activation of a layer, $X_o$, is obtained by multiplying the input activation vector with the weight matrix followed by addition of a bias term, and finally passing the result through an non-linear function such as ReLU,  $X_o=ReLu(X_i*W + b)$. It is evident from this equation that the dominant operation in NN is Multiply \& Accumulate (MAC) in matrix multiplications and bias term additions, of which multiplication is the most hardware intensive. For a fully connected NN with input layer of size $N_{input}$, $K_h$ hidden layers each of size $N_{hidden}$, and output layer of size $N_{output}$, the total number of required multiplications to classify a single sample is, $(N_{input}* N_{hidden})+(N_{hidden}*N_{hidden})*(K_h-1)+ (N_{hidden}* N_{output})$.

Due to their robustness and accuracy, deep CNNs have become the standard for image and pattern recognition \cite{AA1}\cite{CNN}. The operation of a deep CNN is briefly shown in Fig. \ref{fig:cn}(b). During training and inference, each image or pattern is convolved successively with a set of filters where each filter has a set of kernels. After ReLU activation and pooling, the convolution operation is repeated with a new set of filters. Finally, before the output stage, fully connected NNs are used. The convolution operation is shown in Fig. \ref{fig:cn}(c) and it consists of dot products between the input feature-maps and filter weights, mathematically, $f_{out}(m,n)=\sum_{j}\sum_{k} h(j,k)f_{in}(m-j,n-k)$. For a single convolution layer, the total number of multiplications is given by, $N_{in\_channel}*D_k*D_k*D_f*D_f*N_{out\_channel}$, where $D_k$ is kernel dimension, $D_f$ is output feature map dimension, $N_{in\_channel}$ is number of input channels, and number of output channels is $N_{out\_channel}$. For deep CNNs, the total number required multiplications to classify each input image/pattern is a substantial number, and MAC operations account for 90\% or more of the total computational cost \cite {AA1}-\cite {AA4}.

\begin{figure}[h]
\centering
\vspace{-0.05in}
\includegraphics[width=2.9in,height=1.6in]{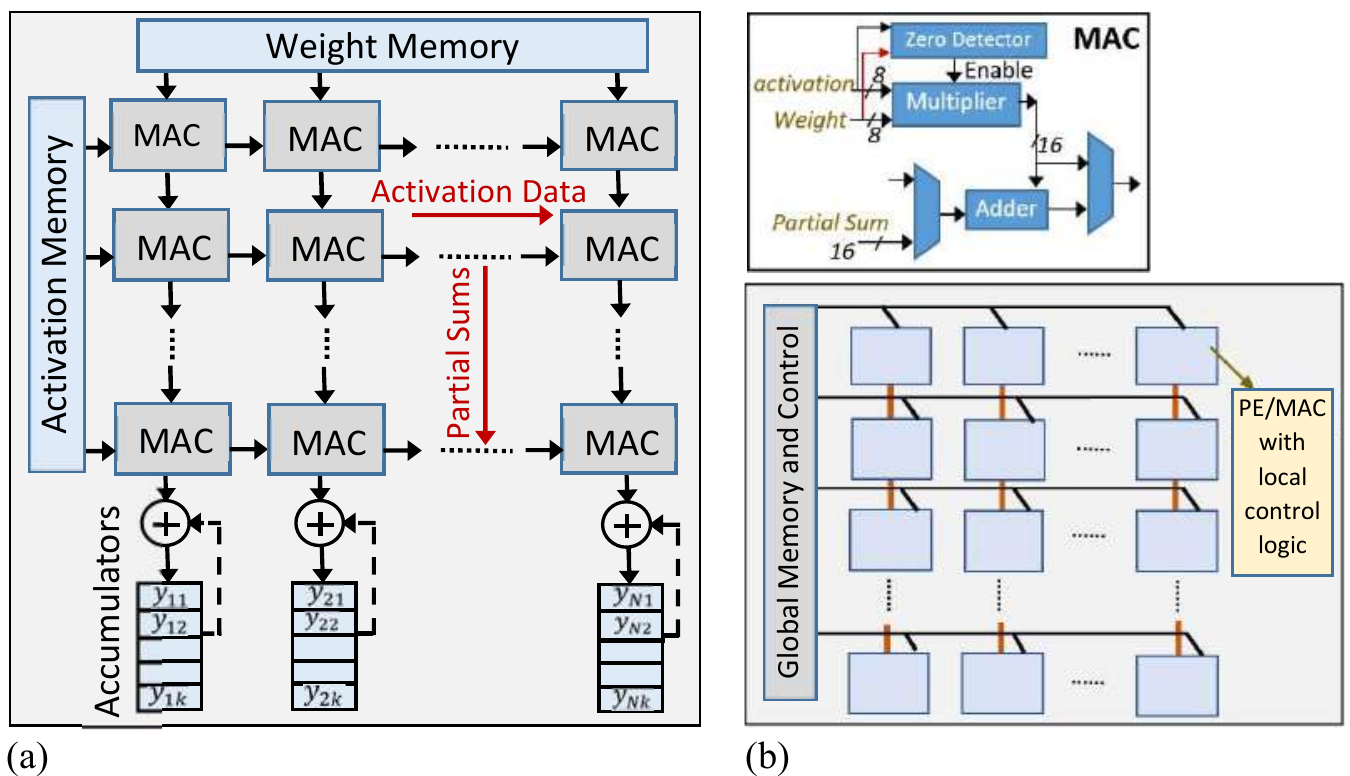}
\vspace{-0.05in}
\caption {AI accelerator with PE/MAC arrays. (a) Systolic architecure; (b) SIMD architecture.}
\label{fig:PE}
\end{figure}

Since the computations in NN/CNN are mostly dominated by MAC operations, the AI accelerators are primarily occupied with arrays of Processing Elements (PE) optimized for fast MAC function \cite{AA1}. Depending on if the accelerators are for training and inference or only for inference on edge devices, the MAC units can be of 32 bits supporting FP32 or 8 bits for int8 \cite{int8} quantized operations \cite{AA1}\cite{QQ}. As shown in Fig. \ref{fig:PE}, the accelerator architectures can be categorized into two domains,  (i) tightly-coupled 2D systolic-arrays (e.g., Google’s TPU) \cite{AA4}\cite{gpu}, and (ii) loosely coupled spatial arrays with independent PEs interconnected with NoC/mesh and using  SIMD architecture \cite{AA2}\cite{sim_sys}. While spatial SIMD designs offer higher flexibility by allowing independent control and deactivation of PEs as needed, systolic arrays must work in lock-step and need more sophisticated techniques to deactivate individual PEs \cite{NY}\cite{sys_f}.

 To optimize silicon utilization, the large number of PE/MACs are densely placed on the chip die \cite{AA1}-\cite{AA4}. As a result, the yield of the accelerator will be primarily dictated by circuit faults in MAC modules of the PEs. Categorization of the location and rate of the faults as critical or non-critical with respect to the desired accuracy of the AI task, and corresponding fault-criticality-aware test pattern generation can save costs associated with final yield evaluation. Also, this will allow more chips to pass the functional test (within an accuracy bound of the AI task) resulting in improved yield (i.e., more accelerator passing the quality test at different  specs or product bins).

\section{YAOTA: Yield  and Accuracy Aware Optimum Test for AI Accelerators}

\label{sec:fault_accuracy}
The semiconductor manufacturing defects and corresponding circuit faults in the AI accelerator need a deeper investigation with regards to its impact on the accuracy of AI and deep learning workloads.  The key modules of accelerators that are susceptible to defects are the weight storage SRAM/Register Files (RF) and the MACs. For SRAM/RF, the faults are generally repaired with ECC and spare cells. The timing-faults in MAC can be solved by appropriate timing guard-band and run-time frequency adjustment. The stuck-at faults in the MAC are permanent and severe, and thus in this work, we focus on stuck-at faults. Any stuck-at faults on the bfloat16/float32 format \cite{bfloat_2} or 8-bit (for int8 quantization \cite{int8}) multipliers and adders will cause a certain precision loss and inaccuracy at the output of MAC. The important question is, \textit{``in an AI accelerator with thousands of PEs with MAC units, will the presence of a few faulty MACs cause the whole accelerator chip to be discarded, resulting in the loss of yield and revenue?"} A scientifically pragmatic solution would be to assess the impact of MAC circuit faults on the training and inference accuracy of AI workloads executed on these accelerators. The key factors to consider in this assessment are, (i) location of the faults inside the PE and its impact on the precision of MAC output, (ii) the fraction of the total PEs that have faulty MACs, (iii) the type of AI workload, and (iv) if the accelerator is for both training and inference, or inference-only.

\subsection{Fault location, MAC Precision and AI Accuracy}
Systematic defects and yield losses are caused by layout-sensitive lithographic hotspots and other process imperfections, variations, and are generally independent of the layout area \cite{Y1}. On the other hand, random defect generated yield losses are caused by defect particles and are dependent on the standard-cell or the layout area as well as the defect particle size \cite{Y4}\cite{Y3}. These defects (e.g., short/open defect, poor contact/via, etc.) and corresponding circuit faults can occur at different sites inside the MAC circuit block. The precision loss - due to the presence of circuit faults - at the output of multiply and accumulate operation will depend on the location of the fault inside MAC circuit and the logic cone impacted by the fault. For example, if a multiplier circuit that performs the multiplication of two 8-bit numbers, has faults impacting upto $K$ LSB bits, then it will sustain worst-case error of $\pm\sum_{i=0}^{K+1}2^i$ (the last $2^{K+1}$ term comes from worst-case carry-in path of partial product addition, as explained in the next subsection). As $K$ increases, the faults impact the more significant digits causing the worst-case error of the multiplier to increase. Similarly, errors will also occur in the adder circuit of the MAC if it is corrupted by faults. Since the multiplier is the more dominant block in a MAC, it will be more prone to faults and computation errors.

As explained in Section II(B), the computations in NNs and CNNs for an AI workload execution are heavily dominated by the MAC operations. Any inaccuracy in MAC output will impact the accuracy and efficacy of the AI task running on the accelerator. From the matrix multiplications in NN and convolution operations in CNN, it can be inferred that despite a certain amount of error in a few MAC modules, the AI tasks can be accomplished with minimum accuracy loss. The relationship between worst-case error per faulty MAC, percentage of MACs that are faulty and the AI workload's accuracy loss due to these errors are correlated, and will depend on the specifics of the NN/CNN architecture and the workload as will be demonstrated in Section IV.

\subsection{Yield and Accuracy Aware Fault Isolation and Test Pattern Generation}

\begin{figure}[h]
\centering
\includegraphics[width=2.75in,height=1.1in]{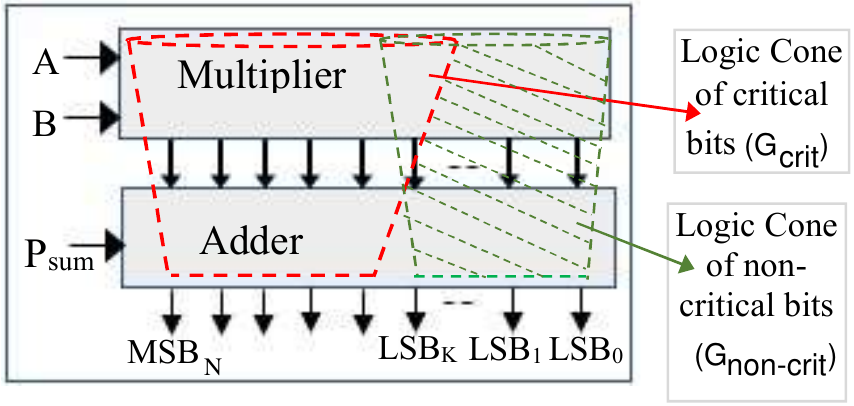}
\caption {Simplified block diagram of MAC unit with logic cones.}
\label{fig:cone}
\end{figure}

\begin{algorithm}
\caption{Isolate critical and non-critical faults, and generate test patterns accordingly for Multiplier and Adder}
\scriptsize
\begin{algorithmic}[1]
\Procedure{Generate test patterns for critical and non-critical faults}{}\\
\textbf{Input:} Gate-level netlist of the MAC module \\
\textbf{Input:} Maximum LSB bit position of MAC output that can tolerate errors, $K$  \\

\textbf{Output:} List of critical faults, $F_{crit}$  \\
\textbf{Output:} List of non-critical faults, $F_{non-crit}$  \\
\textbf{Output:} ATPG test patterns for critical faults, $T_{crit}$  \\
\textbf{Output:} ATPG test patterns for non-critical faults, $T_{non-crit}$  

\State Initialization: , $G_1=\{\}$
\State Initialization: , $G_2=\{\}$

\For {$i$ = $0$ to max. LSB position $K$ }
\State $g_{i} \gets$all gates in the fan-in logic cone of output LSB of position $i$ 
\State $G_{1}\gets G_{1} \cup g_{i}$
\EndFor 

\For {$i$ = $K+1$ to MSB position $N$ }
\State $h_{i} \gets$all gates in the fan-in logic cone of output bit of position $i$ 
\State $G_{2}\gets G_{2} \cup h_{i}$
\EndFor 

\State $G_{non-crit} = G_{1} - G_{2}$

\State $G_{carryin\_bit_{K+1}}\gets$all gates in the logic cone of carry\_in pin of output bit $K+1$

\State $G_{crit} = G_{2} - G_{carryin\_bit_{K+1}}$

\State $F_{crit} \gets $list of all faults (e.g., stuck-at)  for gates in $G_{crit}$

\State $F_{non-crit} \gets $list of all faults (e.g., stuck-at)  for gates in $G_{non-crit}$

\State $T_{crit} \gets $ATPG tool generated test patterns for faults in list $F_{crit}$
\State $T_{non-crit} \gets $ATPG tool generated test patterns for faults in list $F_{non-crit}$

\EndProcedure 
\end{algorithmic}
\end{algorithm}

The precision loss and the extent of computational error in a MAC will depend on the output bit positions that were corrupted by the circuit faults (e.g., stuck-at and delay). In this paper, we focus on stuck-at faults as they are more destructive compared to delay faults. By analyzing the fan-in logic cone of an output bit, we can isolate the circuit paths and standard-cell logic gates that can contribute to a stuck-at fault at that output bit. This can be explained with the multiplier and adder circuit of Fig. 3, and Algorithm 1. In Fig. 3 the green (shaded) logic cone consists of all the logic gates that are \textcolor{black} {exclusively located (i.e, not overlapped with the fan-in cone of the rest of the bits)} in the fan-in cone of the first $K$ Least Significant Bits (LSB) of the output . The red cone contains the standard-cells that are present in the fan-in logic cone of output bits at positions $K+1$ to the MSB bit $N$, where $N>K$. From an extensive execution of AI benchmarks, we identify the acceptable error range of a faulty MAC and the fault rate that will not cause the AI task's accuracy loss to exceed a threshold. We define output bit up to $K$ LSBs to be non-critical from this workload-driven analysis. If circuit faults located within the logic cone of the first $K$ output bits are non-critical then the resulting worst-case error of the MAC is $\pm\sum_{i=0}^{K+1}2^i$, because each bit position $i$ - depending on whether stuck-at-1 or stuck-at-0 - will introduce error $\pm2^i$, and an additional worst-case error of $\pm2^{K+1}$ may occur because of possibly wrong carry propagation to the critical output bit position $K+1$ from non-critical bit $K$. The flow to isolate the critical and non-critical faults of a MAC for AI tasks, and corresponding ATPG pattern generation is presented in Algorithm 1. The inputs to the algorithm are the gate-level netlist of the MAC and the bit position $K$ up to which the error is acceptable. The outputs  of Algorithm 1 in Lines 4 to 7, are the lists of critical ($F_{crit}$) and non-critical ($F_{non-crit}$) faults, and the ATPG patterns ($T_{crit}$, $T_{non-crit}$) that can detect these faults. In Lines 10 to 13, for each bit position from 0 to $K$, the standard-cell logic gates in the logic cone of that bit are identified and added to the list $G_1$. Similarly, in Lines 14 to 17 the logic gates in the fan-in cone of the rest of the bits, $K+1$ to $N$ (MSB), are obtained in the list $G_2$. In Line 18, the list $G_2$ is subtracted from $G_1$ to obtain the list of non-critical gate list $G_{non-crit}$, \textcolor{black} {thus removing any overlap with rest of the critical group}. In Lines 18 to 19, the gates in the carry-in path of bit $K+1$ are identified and subtracted from $G_2$ to obtain the list of error critical gates $G_{crit}$. Finally, in Lines 21 to 24, the fault lists for these gates - $F_{crit}$  and $F_{non-crit}$ - are fed to the ATPG tool to obtain the test pattern sets $T_{crit}$ and $T_{non-crit}$ to test the AI-accuracy critical and non-critical faults, respectively.

\textcolor{black} {In this logic cone analysis, we assume ripple-carry adders are not used in the MAC module as the width of the adder/multiplier required for the AI tasks are at least 8 bits, and ripple-carry structures in these cases will cause long critical paths and hence slower speeds. Our analysis assumes the use of faster carry-lookahead or tree adders (the standard type used in 8 bit or wider cases) in the MAC. Note that, in the unusual event that a slower ripple-carry structure is used, the logic cone of LSB bits will completely overlap with the rest of the bits. However, still in this scenario, our above method of non-critical gate identification will hold if we break the ripple carry path after $K$ LSB bits, and it will incur max error of $\pm\sum_{i=0}^{K+1}2^i$ as discussed above.}

\begin{figure}[h]
\centering
\includegraphics[width=2.83in,height=2.7in]{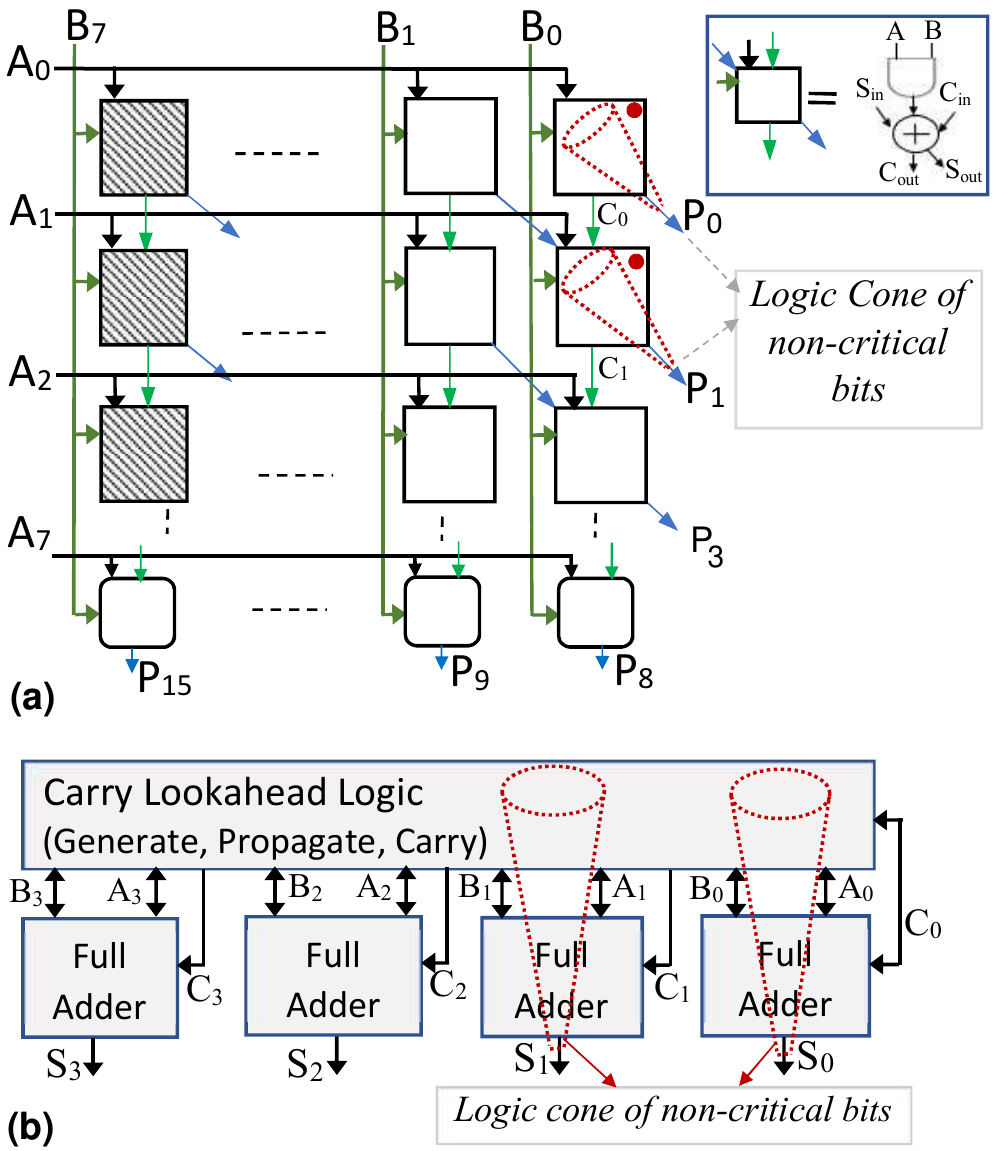}
\caption {Int8 \cite{int8} multiplication and addition. (a) 8-bit Baugh-Wooly signed multiplier (logic cones of two LSBs are shown); (b) first 4 bits of the 16-bit CLA adder  (logic cones of two LSBs are shown.)}
\label{fig:delay_r}
\end{figure}

\begin{figure}[h]
\centering
\includegraphics[width=3.5in,height=1.3in]{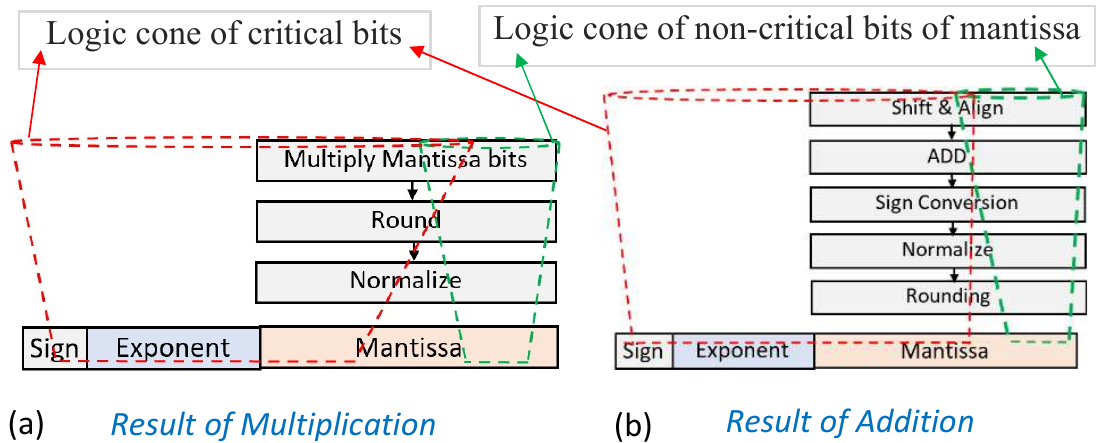}
\caption {Logic cones of critical and non-critical bits are shown for bfloat16/float32, (a) floating point multiplication; (b) floating point addition.}
\label{fig:delay_r}
\end{figure}

The standard data type used in the AI domain to represent the NN connections, CNN filter weights, and activation outputs are the 8-bit int8 quantized format or bfloat16/float32 format \cite{AA1}\cite{AA4}\cite{QQ}\cite{bfloat_2}. Bfloat16/float32 can be used for both training and inference, but for inference-only devices (e.g., mobile and other edge devices) the training is generally done in Bfloat16/float32 \cite{bfloat_1}\cite{bfloat_2} and then the learned parameters are quantized to int8 and loaded in these devices \cite{AA3}\cite{QQ}. The Baugh-Wooly multiplier \cite{DW} - widely used for multiplying two 8-bit signed numbers - is shown in Fig. 4 (a). If acceptable faults can only  affect the two LSBs (i.e., $K=1$ in Algorithm 1), all circuits in logic-cones of outputs $P_0$ and $P_1$ - highlighted with red dots in Fig. 4 (a) - are considered non-critical, and the rest of the circuits are fault-critical. Similarly, in Fig. 4 (b) for the 16-bit Carry Look-ahead Adder (CLA) of the MAC, the logic cones containing the non-critical circuits and faults are shown for $K=1$. For wide-bit addition (i.e. 16-bit), CLA or tree adder are used instead of the slower ripple carry structure \cite{DW}. For AI accelerators using bfloat16/float32 format multiplier and adder \cite{bfloat_1}\cite{bfloat_2}, as shown in Fig. 5, circuit faults in logic cones of some of the lower order (i.e., LSBs) bits of mantissa can be considered as non-critical, and the circuit faults present in the logic cones of the rest of the bits of mantissa, the exponent and sign bits are considered critical as they can introduce large errors in the result. The number of bits in mantissa that can be considered non-critical will depend on the amount of error introduced by stuck-at faults on those bits and the impact of this error in AI task's accuracy loss.

Next, we propose the test flow to identify faulty PEs, and the use of a low area-overhead register memory to store the IDs of the faulty PEs. A control scheme is presented to deactivate some of the faulty PEs during inference and test, if required.

\begin{figure}[h]
\centering
\includegraphics[width=2in,height=1.4in]{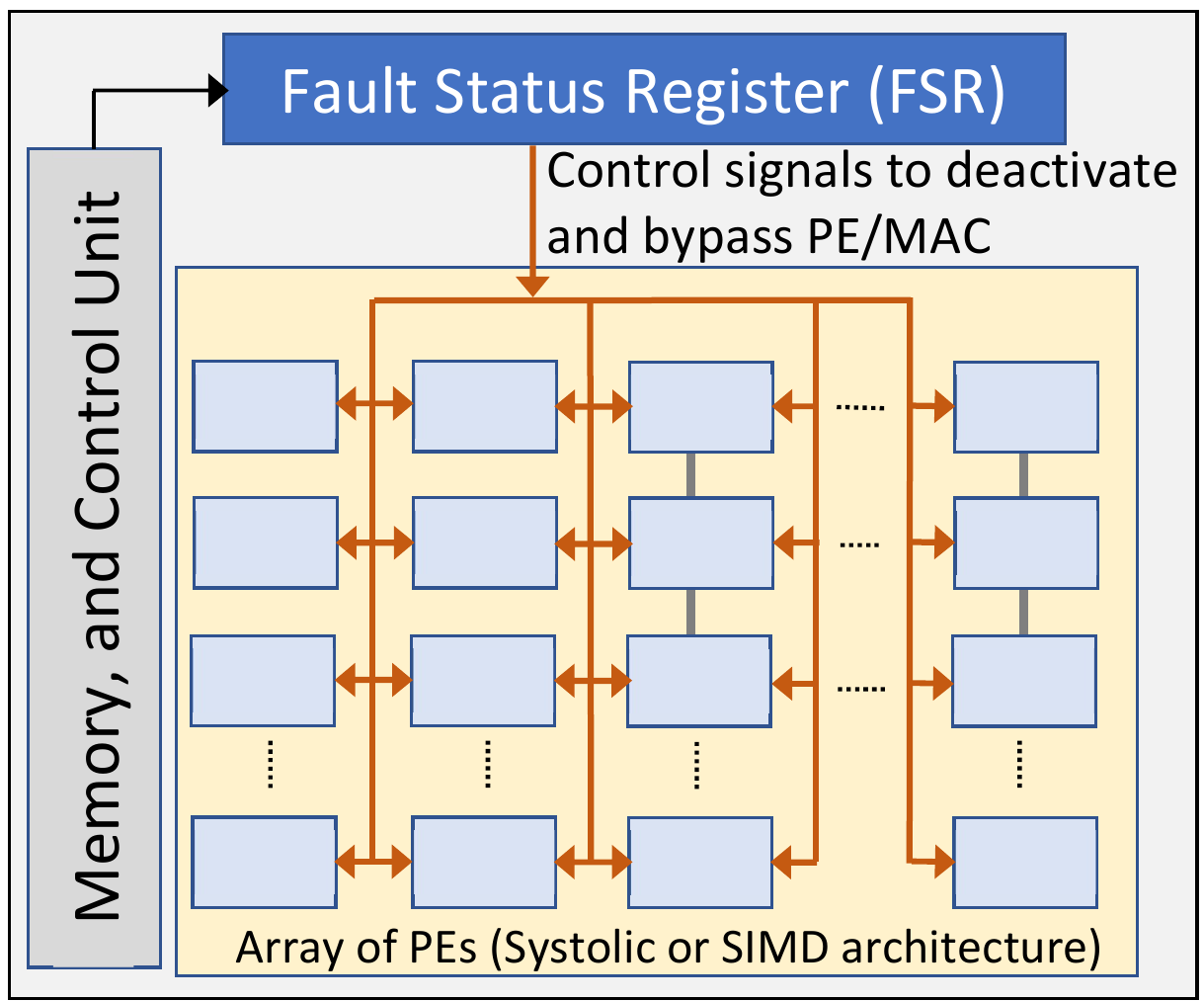}
\caption {Proposed Failure Status Register (FSR) and controls to deactivate and bypass faulty PEs in the accelerator as needed.}
\label{fig:acc}
\end{figure}

\subsection{Post-fab Test of the AI Accelerator} 
After manufacturing, the accelerator chip/die is subjected to structural and functional tests to identify if the PEs in the densely packed arrays are functionally correct. Because of defects from imperfections in the semiconductor manufacturing process, some of the PEs will have faulty MAC units. First, the test patterns $T_{crit}$ obtained using Algorithm 1, are applied in a broadcast mode to all the PE/MACs to identify which units will introduce large errors during matrix multiplications or dot products in convolutions. The IDs of these faulty PEs are recorded in the list $Fail\_ID_{crit}$. Next, the test patterns $T_{non-crit}$ are applied parallelly to all the PEs that passed the first test. The IDs of the PEs that failed this second test are recorded in the list $Fail\_ID_{non-crit}$. Since the PEs that belong to $Fail\_ID_{crit}$ have large MAC errors, they can be permanently disabled by the manufacturer (similar to the practice of disabling faulty cores in many-core processors \cite{core}), or their $Fail\_ID_{crit}$ data can be written to an on-chip non-volatile memory - Fault Status Register (FSR) shown in Fig. \ref{fig:acc} - for the customer to decide if they want to use those PEs or not. The PEs belonging to  $Fail\_ID_{non-crit}$ have relatively small MAC errors and can be acceptable depending on, (i) how many such errors exist (i.e., number of elements in $Fail\_ID_{non-crit}$), (ii) the type of AI workload that will be executed by the customer and their accuracy tolerance limits. Hence, the $Fail\_ID_{non-crit}$ contents are written into the on-chip FSR for the customer to disable a fraction of these faulty PEs with software at runtime and still accomplish the AI tasks with minimal accuracy loss. \textcolor{black}{The deactivation protocol and the complete map of faulty PE locations are programmed in firmware/software by the manufacturer. With the user-given input of acceptable fault rate (in non-critical LSBs, analyzed in Section IV) and the stored PE fault map, the protocol will automatically disable a few faulty PEs  to ensure that the overall faulty PE rate of the accelerator does not exceed a threshold. When deactivating some faulty PEs, the firmware/software will ensure that the remaining faulty PEs are not clustered. We define the deactivation protocol such that after deactivation, the remaining faulty PEs (with non-critical faults) are uniformly distributed across the columns.} As shown in Fig. \ref{fig:acc}, control signals are transmitted from the FSR to all the PEs to selectively disable the faulty PEs when needed. By adopting this scheme, the manufacturer can avoid discarding the full accelerator chip/die only because of the presence of few PEs with faulty MACs, and thereby increase yield. The overhead in this yield loss reduction are the extra on-chip register (FSR) to store the IDs and the control signal routes to disable faulty PEs.

\begin{figure}[h]
\centering
\includegraphics[width=3.1in,height=1.8in]{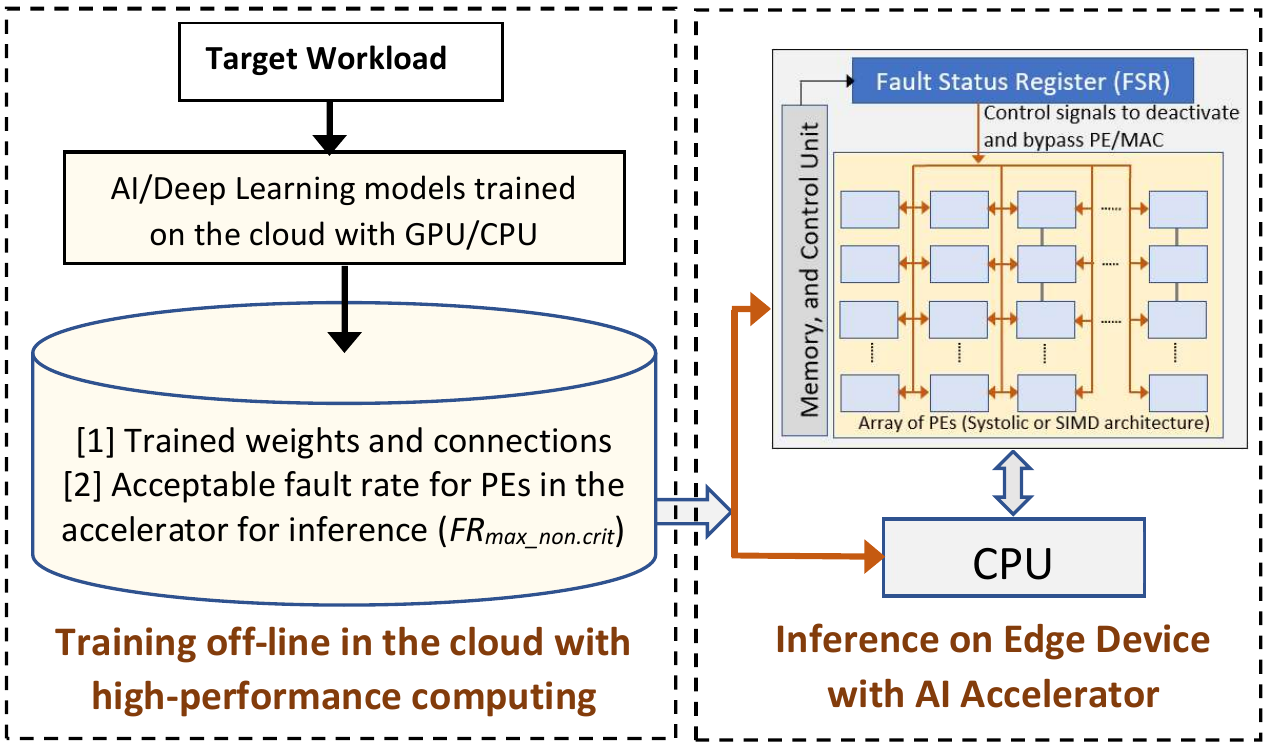}
\vspace{-0.05in}
\caption {Accelerator used for Inference on edge/mobile devices with cloud-trained parameters.}
\vspace{-0.05in}
\label{fig:delay_r}
\end{figure}

\subsubsection{Accelerator used for Inference Only}
For large datasets, the training of deep NN/CNN is computationally intensive and very time consuming,  and generally requires many accelerators or CPU/GPUs working in parallel. For example, training the popular CNN model of AlexNet on imageNet \cite{img} dataset required 6 days with 2 GPUs \cite{Time}.  Since mobile and other edge devices cannot sustain this computational overhead, the general trend for these devices is to use a pre-trained model during inference. In this mode, using many CPU/GPUs and dedicated accelerators the AI model is trained on the cloud, and after training the model parameters and weights are loaded in the edge device where a local AI accelerator is used to perform the MAC operations required for inference \cite{AA1}\cite{QQ}. Our proposed ‘fault-aware cloud-trained edge-inferred’ inference flow is shown in Fig. 7. In this approach, after the model has been completely trained in the cloud with high-performance computing, an additional analysis is performed to obtain the impact of MAC errors on inference accuracy. This can be achieved by injecting on the post-trained model in the cloud the same MAC errors that would occur in the inference accelerator of the edge device, and obtaining the corresponding accuracy changes as a look-up table of fault rate vs. accuracy changes. In summary, the proposed flow will not only generate the trained parameters of the AI model - similar to the regular cloud-training and edge-inference paradigm \cite{AA1} - but also report the maximum allowed fault rate $FR_{max\_non-crit}$ to be subsequently loaded into the accelerator hardware of the edge device as a bitstream. 
During inference, the mobile/edge accelerator's FSR and control unit reads the $FR_{max\_non-crit}$ and if it is lower than the current fault rate $FR_{non-crit}$, then the control unit sends deactivation signals to disable some of the PEs such that the fault rate is reduced to $FR_{max\_non-crit}$ as shown in Fig. 7.           

\begin{figure}[h]
\centering
\includegraphics[width=3.1in,height=1.9in]{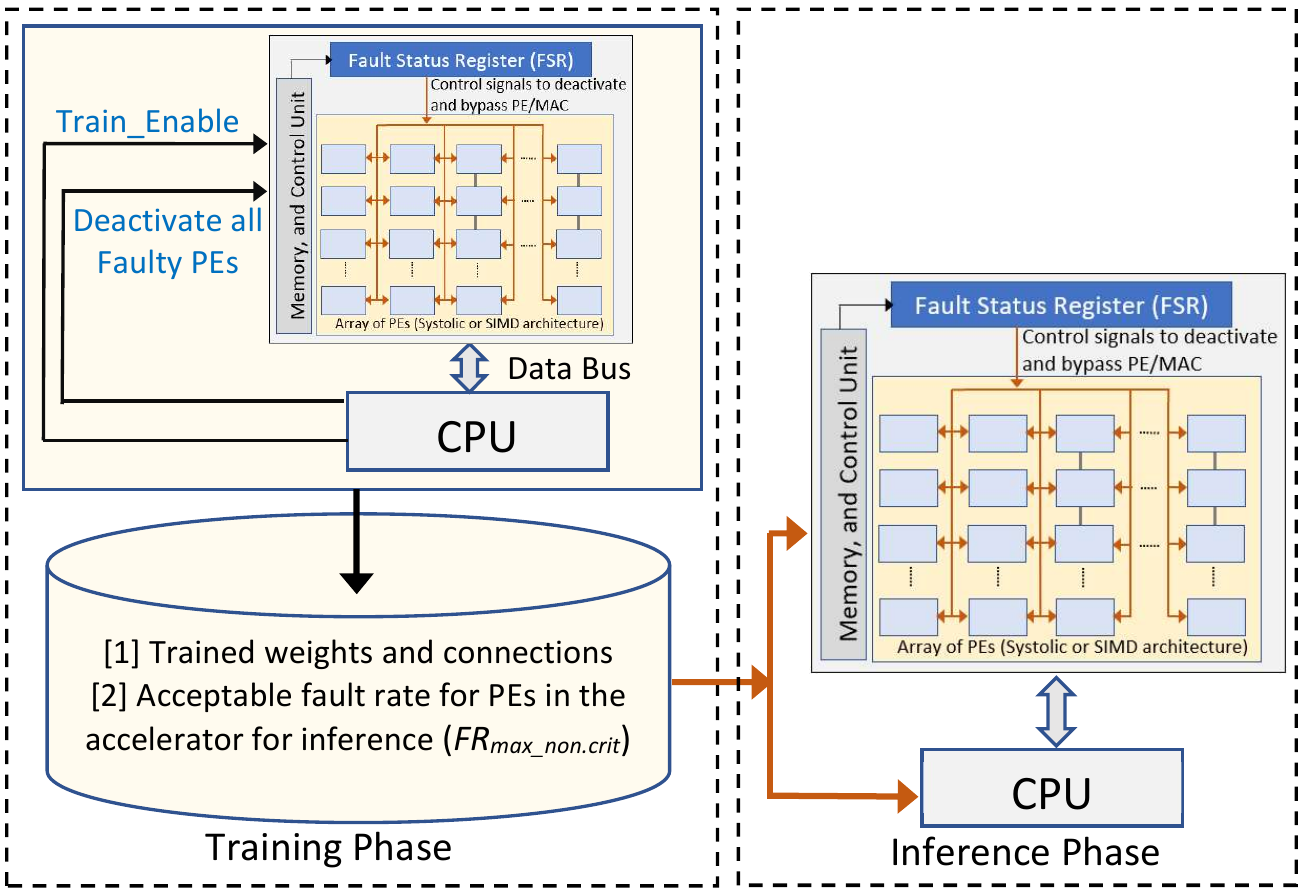}
\vspace{-0.05in}
\caption {Accelerator used in both Training and Inference.}
\label{fig:delay_r}
\end{figure}

\begin{algorithm}
\caption{\small Fault-aware training and inference on AI accelerator}
\scriptsize
\begin{algorithmic}[1]
\Procedure{Identify the maximum acceptable fault rate of Processing Elements (PE) in the accelerator for a given accuracy target}{}\\
\textbf{Input:} Minimum acceptable accuracy for inference, $Acc\_Inf\_threshold$  \\
\textbf{Input:} Training dataset for the deep CNN/NN \\
\textbf{Input:} Fault rate (i.e., rate of non-critical faulty PEs ) of the accelarator, $FR_{non-crit}$ \\

\textbf{Input:} IDs of all faulty PEs in the accelerator, $F\_ID=\{Fail\_ID_{non-crit},Fail\_ID_{crit}\}$  \\

\textbf{Input:} Fault rate adjustment step size, $\delta_{step}$ \\
\textbf{Output:} Maximum allowed fault rate in the AI accelerator for inference at desired accuracy, $FR_{max\_non-crit}$
\State Initialization: Training accuracy, $Acc\_Train=0$
\State Initialization: Deactivate all faulty PEs in $F\_ID$  during training mode
\State Initialization: $FR_{max\_non-crit}=FR_{non-crit}$ 
\While {($Acc\_Train < Acc\_Inf\_threshold$)}
\State $Acc\_Train\gets$ training accuracy for the given dataset with $FR_{max\_non-crit}$ modeled in the backpropagation 
\If {($Acc\_Train < Acc\_Inf\_threshold$)}
\State $FR_{max\_non-crit}=FR_{max\_non-crit}-\delta_{step}$
\EndIf 

\EndWhile
\EndProcedure 
\end{algorithmic}
\end{algorithm}

\subsubsection{Accelerator used in Training and Inference}For cases where the accelerator will be used for both training and inference, further improvement in accuracy degradation caused by MAC errors can be accomplished with fault-aware training. Retraining is a popular tool in deep learning, and primarily used to reduce the size of the NN/CNN by pruning \cite{P1}-\cite{P3}. In retraining, the forward pass and backward passes are made aware of the changes in the network, and this directs Stochastic Gradient Descent based backpropagation flow to evolve the weights of the NN/CNN accordingly to minimize any accuracy loss stemming from these changes \cite{P2}\cite{A1}\cite{A2}. The proposed methodology is shown in Fig. 8 and explained in Algorithm 2. The presence of possible errors in MAC calculations during inference is modeled mathematically and fed to the backpropagation weight-update flows similar to \cite{A1}-\cite{A4}. During the training phase, first, the control unit reads the FSR to obtain the IDs of faulty PEs. To ensure that the CNN/NN will be trained to achieve the best accuracy, all faulty PEs (both critical and non-critical) are disabled during the training phase as shown in Fig. 8 and Line 9 in Algorithm 2. In Algorithm 2 the fault-aware AI training and inference flow is shown for the accelerator. The minimum acceptable inference accuracy ($Acc\_Inf\_threshold$), the training data set, accelerator's non-critical fault rate ($FR_{non-crit}$), IDs of all faulty PEs and a fault adjustment step size ($\delta_{step}$) are provided as inputs to the algorithm in Lines 2 to 6. The algorithm reports the maximum allowed faulty (non-criticalin LSBs) PE rate (i.e., the fraction of the total PEs in the accelerator that have non-critical faults), $FR_{max\_non-crit}$ that will allow the achievement of desired inference accuracy $Acc\_Inf\_threshold$. In Lines 11 to 16, after each iteration of fault-aware training - with fault effect modeled in the backpropagation - the obtained accuracy is compared with the desired inference accuracy $Acc\_Inf\_threshold$. If the accuracy goal is not met, the fault rate $FR_{max\_non-crit}$ is reduced by a small step $\delta_{step}$, until the desired accuracy goal is met. After the training converges and the inference accuracy target is reached, the trained weights of the NN/CNN are obtained. Additionally, the maximum allowed faulty PE rate, $ FR_{max\_non-crit}$, is reported. During inference, the control unit and FSR will read this $ FR_{max\_non-crit}$ and disable some faulty PE to achieve this $ FR_{max\_non-crit}$ rate.

The benefits from this combined training-inference methodology are, (i) in achieving the same inference accuracy, the fault-aware training approach will allow the accelerator to endure a higher $FR_{max\_non-crit}$ compared to the scenario where no fault-aware training is used. This will allow the availability of more PEs during inference if fault-aware training was used. (ii) Although there will be additional time required for this training step, this extra cost will be amortized on the multiple inference runs. Because, the general trend in AI is to train the NN/CNN once accurately, and then this pre-trained model is used many times during inference. As a result, the extra PEs - made available by the fault-aware training - present during each inference will significantly speed up the inference tasks.

\begin{figure}[h]
\centering
\includegraphics[width=2.3in,height=3.1in]{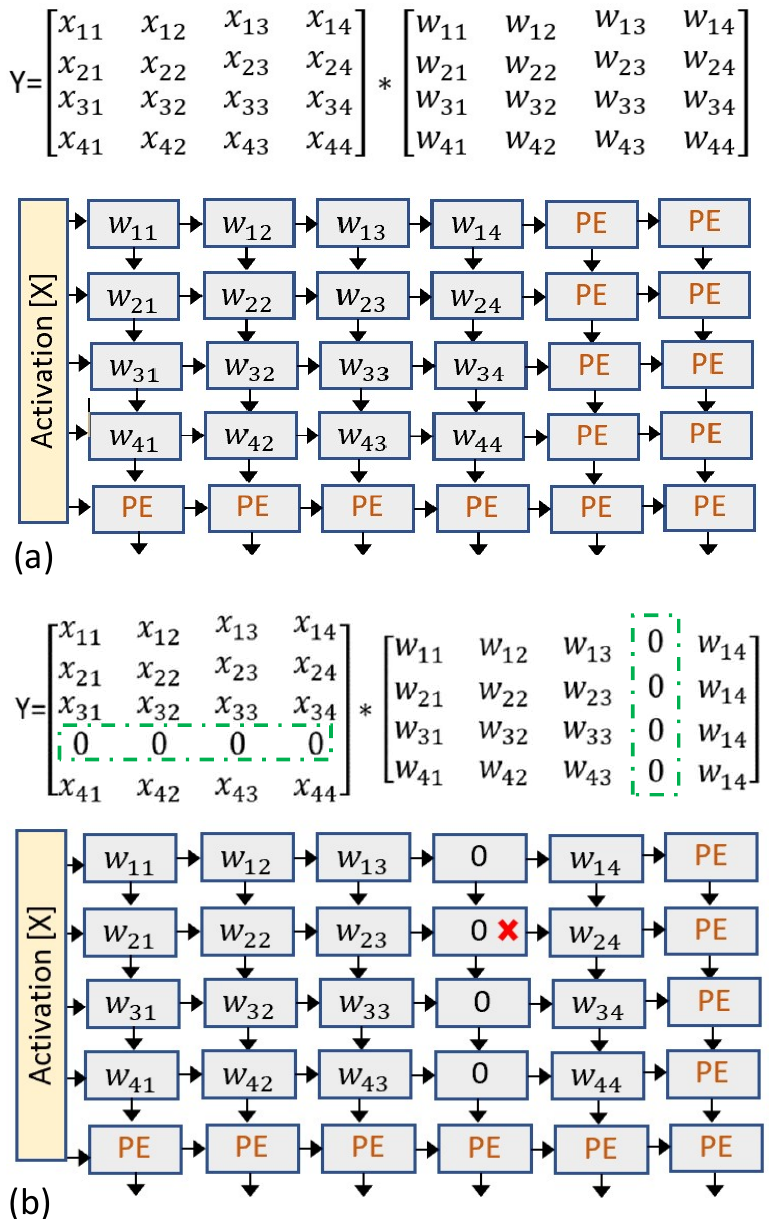}
\vspace{-0.05in}
\caption {Software-level bypass method for faulty PE in Systolic array. (a) Normal mode without fault. (b) Faulty PE bypassed by shifting input data.}
\label{fig:byp}
\vspace{-0.05in}
\end{figure}

\subsection{Deactivation and Bypass of Faulty PE}
The FSR holds the IDs of faulty PEs. On demand deactivation and bypass of these PEs require different strategies for SIMD and Systolic architectures.
\subsubsection{SIMD Architectures}In SIMD architecture the loosely coupled independent PEs are connected with NoC or mesh and can be individually switched off and bypassed with wires \cite{AA1}\cite{sim_sys}. Using the results from the FSR, all PEs with critical faults and few PEs with non-critical faults are deactivated. \textcolor{black}{If after deactivation, `$N\_{remaining}\_{PE}$' PEs remain out of `$N\_{total}\_{PE}$', then  throughput of the accelerator will be scaled by the factor $(N\_{remaining}\_{PE}/N\_{total}\_{PE})$.}
\subsubsection{Systolic Architectures} The tightly-coupled 2D systolic-arrays in TPU introduce challenges in deactivating and bypassing individual PEs. A common technique is to use spare PE blocks to substitute for faulty PEs \cite{sys_f}. However, this approach requires complex wiring between spare and faulty PEs. In this work, we propose an innovative software-level technique to deactivate and bypass faulty PEs in systolic array without any hardware-level modifications. Unlike \cite{NY}, as discussed in details in Section II(A), this approach does not require the NN/CNN weights that are mapped on the faulty PEs to be pruned (i.e., set to zero). The fault-free scenario is shown in Fig. \ref{fig:byp}(a), where the weights are pre-loaded in the PEs of the systolic array in regular manner. In Fig. \ref{fig:byp}(b), the PE in row 2, column 4 of the systolic array has a critical fault or non-critical fault exceeding acceptable fault-rate (marked with a red cross) and needs to be bypassed. In software, this is accomplished by shifting the column of the weight matrix to the right by inserting a dummy row of zeros. The activation matrix, $X$, is also shifted accordingly as shown in Fig. \ref{fig:byp}(b). Moreover, this software-level approach does not require complex wiring resources as needed in hardware-level spare replacement techniques \cite{sys_f}. \textcolor{black}{However, the execution throughput will decrease as some faulty PEs are deactivated. For the largest weight matrix in CNN/NN multiple iteration steps ($N\_{steps}$) are required in the accelerator array to complete the MAC operations. For example, in AlexNet largest weight matrix is of size 4096 by 9216, and this will require multiple iterations to complete in a systolic array with 256 by 256 PEs (e.g., Google TPU \cite{AA4}). If the row/column size of systolic array is `$N\_{dim}\_{sys}\_{arr}$' and the number of columns in accelerator PE array that has atleast one PE requiring fault-induced deactivation is `$N\_{sys}\_{arr}\_{faulty}\_{cols}$', then software-level deactivation will require extra $(N\_{dim}\_{sys}\_{arr} * N\_{sys}\_{arr}\_{faulty}\_{cols})*N\_{steps}$ MAC operations.}

\section{Results and Analysis}
\label{sec:ExperimentalResults}

\begin{figure}[h!]
	\centering
	\includegraphics[width=3.2in,height=2.2in]{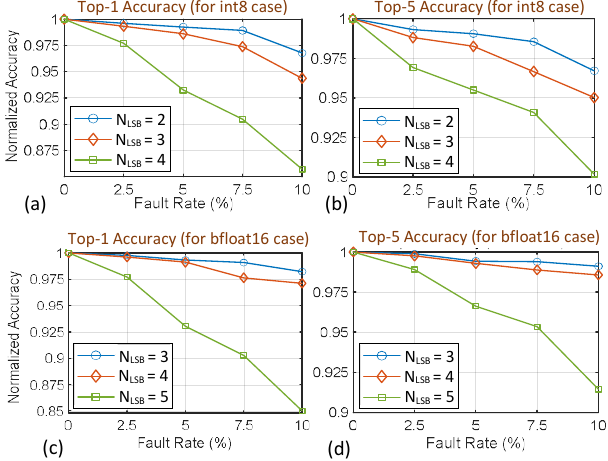}
	\caption {\textcolor{black}{Impact of faults in logic cones of LSB positions  on inference accuracy for AlexNet with ImageNet data set. (a), (b) int8 data format and MAC type.; (c), (d) bfloat16 data format and MAC type.}}
	\label{fig:k_vary}
	\vspace{-0.05in}
\end{figure}

\begin{figure}[h!]
	\centering
	\includegraphics[width=3in,height=1.2in]{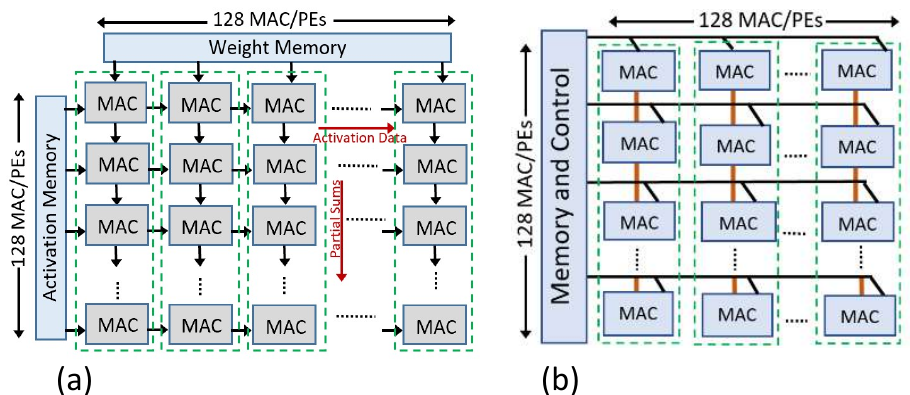}
	\caption {Accelerator with 128x128 MAC/PEs. Faulty MAC/PEs are randomly distributed across the columns. (a) Systolic, (b) SIMD architecture.}
	\label{fig:hm}
		\vspace{-0.05in}
\end{figure}

\textcolor{black}{To identify the impact of the number of LSBs having faults in their logic cones, we varied the number of LSBs from 2 to 4 for int8 \cite{int8} and 3 to 5 for bfloat16 \cite{bfloat_1}\cite{bfloat_2} data formats for the CNN model AlexNet with the ImageNet test set. These experiments were done with PyTorch \cite{pyt}. The results are shown in Fig. \ref{fig:k_vary}, where X-axis is fault-rate in non-critical logic cones (the LSBs). From this analysis, we conservatively selected 2 LSBs for int8 and 4 mantissa LSBs for the recent bfloat16  hardware models.} In our experiments the accelerator hardware comprised of 128 by 128 array of PE/MACs as shown in Fig. \ref{fig:hm}. The faulty PEs are modeled as uniformly distributed across the rows and columns with fault probability of $FR\%$, the fault rate.  This implies that for $FR\%$ fault rate, each column of the accelerator has $0.01*FR*N_{Row}$ faulty PEs randomly distributed across that column. The reported results here are the average of 10 independent fault injection experiments done in MATLAB \cite{mat} (for NN) and PyTorch \cite{pyt} (for CNN) according to this hardware and fault distribution model. \textcolor{black}{The prediction accuracy data reported in subsequent experiments are independent of the SIMD/Systolic architecture of the accelerator. However, the throughput will vary based on architecture type and fault rate as discussed in Section III-D.}  In identifying the critical and non-critical circuit faults in the MAC of a PE and corresponding test pattern generation, we used the RTL of the MAC unit present in each PE and analyzed both int8 and recent bfloat16 format implementations.

For int8 quantized data format, first, the RTL of signed 8-bit MAC unit was developed, followed by gate-level synthesis with SAED \cite{syn} 28nm standard cell library with Synopsys Design Compiler (DC) \cite{syn}. During synthesis, DC tool implemented the signed multiplier using the Baugh-Wooly architecture available in DesignWare \cite{syn}  IP library. The 16-bit adder unit was implemented using the Carry Look Ahead (CLA) structure from the DesignWare IP library. After the complete gate-level netlist of the MAC was available, a custom TCL script was developed for the logic cone analysis of the output bits. From analysis of Fig.\ref{fig:k_vary}, we selected the first two bits of the LSB (i.e., bits 0 and bit 1) as non-critical and  bit positions 2 to 7 as critical, implying that the worst-case total MAC error resulting from these two bits is $\pm(2^0+2^1+2^2$) as explained in Section III(B). Next, using the developed TCL script with DC we obtained critical ($G_{crit}$)  and non-critical ($G_{non-crit}$)  standard cell logic gates present in the logic cones of critical and non-critical bits, respectively,  as described in Algorithm 1 in Section 3(B). Next, the gate-level netlist and the critical and non-critical fault lists were taken to TestMAX ATPG tool \cite{syn}  and three sets of ATPG test patterns were generated – (i) with all faults, (ii) only with the critical faults, (iii) considering only the non-critical faults. The results are shown in Table I. The reason that the number of test patterns for the all-faults case is lower than the sum of critical-only and non-critical-only cases is due to the method of ATPG pattern generation, where a single pattern can sometimes detect faults from both critical and non-critical groups. However, the test pattern counts to test the critical-only faults is less than the all-faults case in Table I. This critical pattern set can be applied first to all the PEs in a broadcast manner to identify if there are any PEs that must be disabled, this is because, having a critical fault in MAC introduces a large magnitude of error. Next, for the PEs that passed the first test, we apply the test patterns from Row 4 of Table I to test the presence of any non-critical faults. If faults are detected by this pattern set, the IDs of the faulty PEs are recorded in FSR memory as explained in Section III(c).

\begin{table}[]
	\centering
\caption {Fault statisitics for signed int8 multiplier and adder}
\begin{tabular}{|c|c|c|c|c|}
\hline
Case                                                                                    & \begin{tabular}[c]{@{}c@{}}Total\\ cells\end{tabular} & \begin{tabular}[c]{@{}c@{}}Number of\\ stuck-at faults\end{tabular} & \begin{tabular}[c]{@{}c@{}}Total test\\ patterns\end{tabular} & \begin{tabular}[c]{@{}c@{}}Test\\ coverage\end{tabular} \\ \hline
\begin{tabular}[c]{@{}c@{}}All faults included\end{tabular}                           & 206                                                   & 878                                                                 & 24                                                            & 100\%                                                   \\ \hline
\begin{tabular}[c]{@{}c@{}}Critical faults\\ only ($F_{crit}$)\end{tabular}          & 198                                                   & 854                                                                 & 23                                                            & 100\%                                                   \\ \hline
\begin{tabular}[c]{@{}c@{}}Non-critical faults\\  only ($F_{non-crit}$)\end{tabular} & 8                                                     & 24                                                                  & 4                                                             & 100\%                                                   \\ \hline
\end{tabular}
\end{table}

\begin{table}[]
\centering
\caption {Fault stats for MAC with bfloat16 Multiplier and FP32 adder}
\begin{tabular}{|c|c|c|c|c|}
\hline
Case                                                                                    & \begin{tabular}[c]{@{}c@{}}Total\\ cells\end{tabular} & \begin{tabular}[c]{@{}c@{}}Number of\\ stuck-at faults\end{tabular} & \begin{tabular}[c]{@{}c@{}}Total test\\ patterns\end{tabular} & \begin{tabular}[c]{@{}c@{}}Test\\ coverage\end{tabular} \\ \hline
\begin{tabular}[c]{@{}c@{}}All faults included\end{tabular}                           & 1669                                                  & 8755                                                               & 217                                                           & 98.74\%                                                 \\ \hline
\begin{tabular}[c]{@{}c@{}}Critical faults\\ only ($F_{crit}$)\end{tabular}          & 1443                                                  & 7552                                                                & 195                                                           & 98.65\%                                                 \\ \hline
\begin{tabular}[c]{@{}c@{}}Non-critical faults\\  only ($F_{non-crit}$)\end{tabular} & 226                                                   & 1188                                                                & 32                                                           & 99.83\%                                                 \\ \hline
\end{tabular}
		\vspace{-0.05in}
\end{table}

\begin{figure}[h!]
	\centering
	\includegraphics[width=3.1in,height=1.15in]{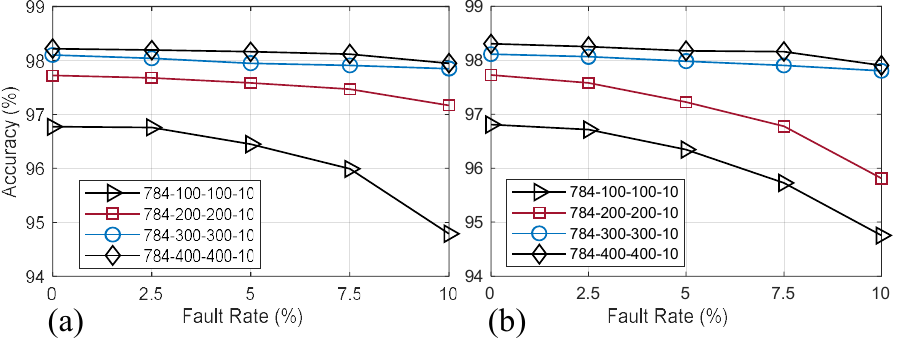}
	
	\caption { Inference accuracy changes with fault \textcolor{black}{(non-critical)} rates for NN running MNIST: (a) Without pruning, (b) With 30\% pruning and retraining }
	\label{fig:nn}
	\vspace{-0.05in}
\end{figure}

As discussed in \cite{bfloat_1}\cite{bfloat_2}, recently, for training NN/CNN bfloat16 method is used where maultiplier is  bfloat16 and accumulator is float32 type. To isolate the critical and non-critical faults of a floating-point MAC, we obtained a floating-point MAC benchmark circuit from OpenCores \cite{oc} and modified it for above mentioned bfloat format. We synthesized a gate-level netlist of the floating point MAC using the DesignWare IP library. For the floating-point MAC, we took the first 4 LSB bits (bits 0 to 5) of the mantissa as non-critical (from analysis in Fig. 10), and the rest of the bits of mantissa, the exponent and sign bit are considered critical. After identifying the critical and non-critical gates and corresponding faults, the fault lists and the gate-level netlist were taken to the ATPG tool and test patterns were generated similar to the int8 case above. The results are shown in Table II. First, the test patterns from Row 3 of Table II are applied to identify all faulty PEs that must be disabled to prevent significant accuracy loss in AI tasks. After that, the non-critical faults are identified with the patterns from Row 4 of Table II. All PEs that failed this second test have non-critical faults and their IDs are recorded in FSR. 
 
To analyze the impact of PE/MAC faults on the inference accuracy of NN, we implemented a 4-layer NN with two hidden layers, and varied the number of neurons in the hidden layers. Also, both un-pruned and 30\% pruned (with-retraining) \cite{P1}-\cite{P3} versions were implemented. For NN experiments we used MATLAB deep learning toolbox \cite{mat}. All weights and activations were quantized in int8 format using MATLAB Fixed-Point tool \cite{mat}. To incorporate the worst-case non-critical MAC faults - obtained from the gate-level netlist above - into the NN inference task, the matrix multiplication function in forward pass of the NN used in inference was modified in MATLAB to inject faults according to the hardware model of Fig. \ref{fig:hm}.   The relationship between accuracy and fault rates are shown in Fig. \ref{fig:nn}. It can be seen from Fig. \ref{fig:nn} (a) that, other than the smaller 100 hidden layer case, the rest of the NNs are robust to faults in the MAC, with normalized accuracy changes less than 0.5\% at 5\% fault rate. With pruning (Fig. \ref{fig:nn} (b)), the accuracy degrades slightly more with faults. This is because with pruning less number of neurons are present, and those that are present become more important.

\begin{table}[]
	\centering
\caption {Multiplication and additions in CNN to classify one image}
\begin{tabular}{|c|c|c|c|c|}
\hline
\begin{tabular}[c]{@{}c@{}}CNN\\ Architecture\end{tabular} & \begin{tabular}[c]{@{}c@{}}Conv2d\\ Layers\end{tabular} & \begin{tabular}[c]{@{}c@{}}Linear\\ Layers\end{tabular} & \begin{tabular}[c]{@{}c@{}}Number of\\ Multiplications\end{tabular} & \begin{tabular}[c]{@{}c@{}}Number of\\ Additions\end{tabular} \\ \hline
LeNet-5                                                    & 3                                                       & 2                                                       & 416,520                                                             & 416,520                                                       \\ \hline
AlexNet                                                    & 5                                                       & 3                                                       & 714,188,480                                                         & 714,188,480                                                   \\ \hline
VGG-16                                                     & 13                                                      & 3                                                       & 15,470,264,320                                                      & 15,470,264,320                                                \\ \hline
ResNet-50                                                  & 53                                                      & 1                                                       & 3,729,522,688                                                       & 1,761,820,672                                                 \\ \hline
\end{tabular}
\end{table}

\begin{figure*}[h!]
	\centering
	\includegraphics[width=7.1in,height=2.55in]{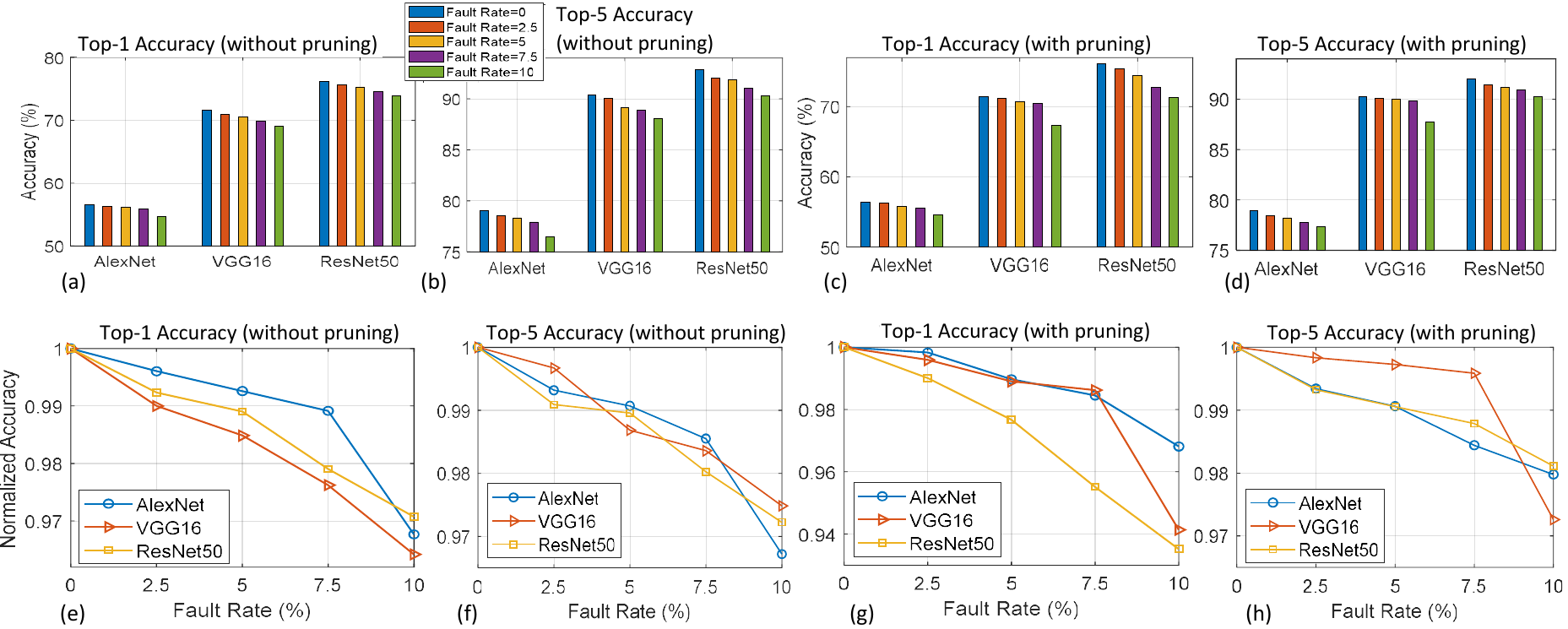}
	\vspace{-0.05in}
	\caption {Inference accuracy changes with fault \textcolor{black}{(non-critical)} rates for CNNs running imagenet dataset. (a), (b) Top-1 and Top-5 accuracy changes ;  (c), (d) Top-1 and Top-5 accuracy changes with model pruning; (e) to (h) normalized accuracy changes in Top-1 and Top-5 (with and without pruning) for fault rates.}
	\label{fig:cnn}
	\vspace{-0.05in}
\end{figure*}

To assess the impact of MAC circuit faults on the accuracy of CNN, we used several key benchmark CNNs – AlexNet \cite{Time}, VGG-16 \cite{vgg}, ResNet-50 \cite{res} and LeNet-5 \cite{CNN}. The number of convolution, linear layers and the total number of multiplication and additions required (without pruning) to classify each image in these networks are tabulated in Table III. These results were obtained using custom functions developed in Pytorch \cite{pyt}. For the smaller CNN, LeNet-5, we performed both training and inference with MNIST dataset \cite{CNN}. For the complex architectures - AlexNet, VGG-16 and ResNet-50 -  training takes several days and requires multiple GPUs \cite{Time}-\cite{vgg}. In Pytorch \cite{pyt} library, pre-trained versions of these CNNs are available where they were already trained with ImageNet \cite{img} dataset having  millions of training images and 1000 possible classes. In our experiments we used these pre-trained models and performed inference with the 50,000 images from the ImageNet \cite{img} validation dataset. During inference, the models were quantized into int8 format using Pytorch’s ‘$torch.nn.quantized$’ library. To incorporate the worst-case non-critical MAC faults - obtained from the gate-level netlist above - in the inference function of the CNN, we used Pytorch’s ‘$register\_forward\_hook$’ feature to access the data in Conv2d function and injected the MAC faults according to the hardware model of Fig. \ref{fig:hm}. The accuracy changes - in the standard Top-1 and Top-5 format - with MAC faults are shown in Fig. \ref{fig:cnn} for 50,000 test images from ImageNet \cite{img}. Top-1 accuracy implies that the predicted class matches exactly the actual class (out of 1000 possible classes) and Top-5 refers to the case where the actual class is within the top 5 predicted classes \cite{Time}-\cite{vgg}\cite{pyt}.  From Fig. \ref{fig:cnn} (e), it can be seen that the normalized accuracy in Top-1 category changes by less than 1.5\% for all networks when fault rates are within 5\%. For the Top-5 category in Fig. \ref{fig:cnn} (f), except for the computationally intensive VGG-16 network, the normalized accuracy degradation was confined within 1\% for fault rates up to 5\%. Next, we pruned 30\% of the filter weights of the convolution layers and repeated our fault injection experiments. Form Fig. \ref{fig:cnn} (g)-(h), it can be seen that, with pruning, the normalized Top-1 accuracy degraded by a small amount with worst-case happening for ResNet-50 where it degraded by 2.2\% for fault rate 5\%. Even with pruning, the Top-5 normalized accuracy degradation was within 1\% for fault rates up to 5\%. Note that, during our pruning experiments on  AlexNet/VGG-16/ResNet-50 retraining was not done, because it would have required us to retrain the networks with 14 million images using a large number of GPUs. Retraining during pruning would improve the accuracy further \cite{P1}-\cite{P3}.

\begin{figure}[h!]
	\centering
	\includegraphics[width=2.2in,height=1.25in]{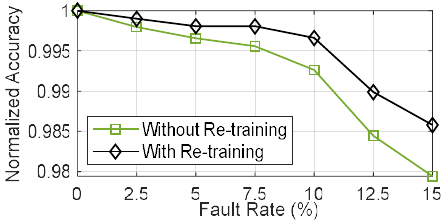}
	\caption {Improvement in accuracy with fault-aware training on LeNet-5.}
	\label{fig:retr}
		\vspace{-0.05in}
\end{figure}

\begin{figure}[h!]
	\centering
	\includegraphics[width=2.2in,height=1.35in]{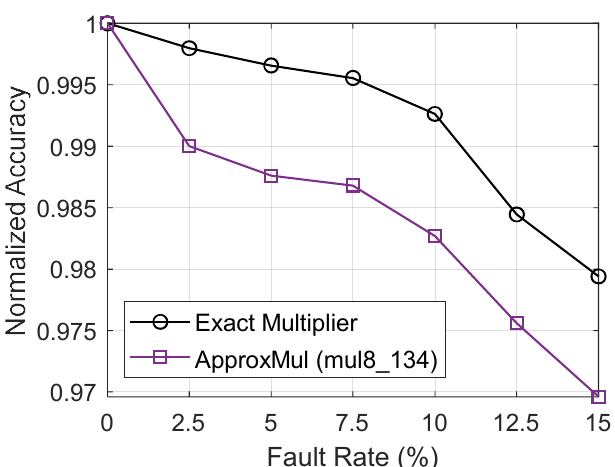}
	\caption {Accuracy change of LeNet-5 CNN with faults for approximate and exact multipliers without retraining. }
	\label{fig:approx}
	\vspace{-0.05in}
\end{figure}

As discussed in Section III(c)(2), using our proposed fault-aware training flow some of the accuracy loss due to faults in MAC units can be recovered by incorporating the fault effects in the backpropagation-based weight update segment and allowing the CNN to adapt accordingly. To experimentally demonstrate this technique, we used the LeNet-5 CNN architecture. We picked the simpler LeNet-5 architecture over AlexNet/VGG-16/ResNet-50 because of the computational complexity of training. Whereas AlexNet/VGG-16/ResNet-50 would require multiple GPUs and several days of training with ImageNet data \cite{Time}-\cite{vgg}, the LeNet-5 can be trained in several minutes on MNIST dataset using CPU. We used 6-core Intel core i7 CPU with 24GB RAM in this training experiment. We modeled the equivalent worst-case MAC error - corresponding to faults occurring in the logic cones of 4 LSB bits of mantissa - in the forward and backpropagation segment using Pytorch’s ‘$register\_hook$’ feature \cite{pyt}. Results from this fault-aware training are shown in Fig. \ref{fig:retr}. \textcolor{black}{The results in Fig. \ref{fig:retr}, corresponds to the bfloat16 format training and inference hardware model as presented in \cite{bfloat_1}\cite{bfloat_2} where multiplier is of bfloat16 type and accumulator is of float32.} It can be seen that for 7.5\% fault rate the normalized accuracy loss improved from 0.5\% to 0.22\% due to fault-aware training.

In Section II(A), we explained that the presence of defect-induced circuit faults in approximate NN/CNN will deteriorate the AI task's accuracy as errors from approximations in MAC were already introduced in the model, and any further error from circuit faults will be  detrimental. In Fig. \ref{fig:approx}, the normalized change of accuracy with respect to faults for LeNet-5 on MNIST dataset is shown for cases of exact  and approximate multipliers.  In \cite{A4} a comprehensive analysis (with approximation aware retraining) of different types approximate multipliers on the efficiency of neural networks were performed, and the best approx. multipliers in terms of prediction accuracy were reported. Based on these findings \cite{A4}, in our experiment as an approx. multiplier, we chose $mul8\_134$ from the open-source library of EvoApprox8b \cite{A5}. As discussed in \cite{A1}\cite{A4}, the accuracy of CNN/NN using approximate multipliers are very sensitive to proper training compared to regular multipliers, and the backpropagation training phase must be updated to account for approximate computing. In our experiment, the initial training phase was updated (using Pytorch's hook functions \cite{pyt}) to account for the use of 8-bit approx. multiplier, also int8 quantization was used. From Fig. \ref{fig:approx}, it can be observed that the presence of faults will degrade the performance of NNs with approx. multipliers significantly. Hence, if yield loss reduction is the primary goal, exact MAC units need to be used to account for possible circuit faults.

From these detailed analyses of gate-level synthesis, fault isolation, ATPG pattern generation for the MAC circuit, and corresponding simulation of fault effects on standard NN/CNN benchmarks, it can be observed that certain circuit faults - based on their locations in the circuit - have minimal impact on the AI task's accuracy when the fault rate is within an upper limit. For example, with 5\% fault rate in non-critical gates, the normalized  Top-5 accuracy loss in CNNs is less than 1\% (Fig. \ref{fig:cnn}(f)(h)). If this 1\% accuracy loss is acceptable, and if there are more than 5\% faulty PEs, then using the IDs of faulty PEs stored in the Fault Status Register, some faulty PEs can de deactivated on each column of the accelerator such that the fault rate is within 5\% on each column of the PE array. For instance, in Fig. \ref{fig:cnn}(h), at 10\% fault-rate the normalized Top-5 accuracy degradation is 3.2\% for AlexNet, but after deactivating few faulty PEs uniformly in each column of the accelerator the fault-rate per column can be reduced to 5\% and this will result in improved Top-5 accuracy degradation to less than 1\%. As a result, an AI accelerator chip with few faulty PEs can be binned accordingly and shipped, improving valuable yield and revenue. The tradeoff in this yield saving would be the lower number of PE blocks in the accelerator due to the deactivation of few faulty PEs to keep the fault rate within an acceptable limit (i.e., 5\%), however, this will not have any functional impact, and will only reduce the throughput marginally. Furthermore, the reduced throughput PEs can be binned and priced differently without totally discarding the chip, thus saving yield. For example, the accelerators with no fault at all can be placed in the top bin and sold at a premium price to be used in safety-critical applications such as self-driving cars, whereas accelerators in the lower bins (with few faults, e.g., less than 5\% fault rate) can be used in other AI/deep-learning tasks that can tolerate errors with minimal performance loss. 

\textcolor{black}{In \cite{NY} when considering faults, authors assumed the complete MAC unit (i.e., all bits) was faulty, in contrast, our approach is more pragmatic and conservative. We analyzed our faults within groups of bits (i.e.,  logic cone of certain LSBs Vs. logic cone of the rest).  If there are any errors beyond a certain LSB position we deactivate that PE to avoid a large extent of error in the accuracy. The complexity and cost of our approach are also minimal. For identifying the allowable non-critical fault rate and the number of LSBs, our search space is limited as we choose few LSB bits and fault rates (Fig. \ref{fig:k_vary}). Moreover, inference is not necessarily time-consuming and if such an analysis is performed once per hardware with an exhaustive benchmark like ImageNet \cite{img}, then it will suffice for any other benchmark. For the fault-aware training part, only the forward pass function needs to be updated to reflect the MAC fault rate that would be present during inference, and this is done for only one fault rate (e.g., 2.5\% or 5\% in Fig. \ref {fig:retr}). Hence, our fault-aware training does not add extra complexity compared to regular training. As we confine our faults within a few LSBs, our approach does not require hardware fault location-aware AI workload mapping and retraining as in \cite{NY}.}

\textcolor{black}{In our experiments we used the state-of-the-art CNNs that were pre-trained with 1 million images from ImageNet \cite{img}. Because of the large training set, these CNNs are very well-trained for any other type of pattern recognition/AI task (i.e., similar to the concept of ``transfer learning'' where ImageNet pre-trained networks can be used for any other pattern recognition problem by adjusting and training the final layers). Also, for validation of our fault effects, we used the 50K sample validation images from ImageNet with PyTorch. As we have shown our approach works on various CNNs trained with ImageNet, it suffices that for any other AI workload the concept is applicable.}

\section{Conclusion}
In this paper, we presented a yield loss reduction and test methodology  for AI accelerator chips densely packed with PEs. Exploiting the error-healing properties of backpropagation during training and the inherent fault tolerance features of trained AI models during inference, we obtained an analytical relationship between fault location and fault rate of MAC, and the AI task's accuracy to guide yield decisions. Simulation results on NN/CNN show that the proposed YAOTA approach will allow up to 5\% faulty PEs in the accelerator at the expense of less than 1\% loss in the AI task's accuracy. Furthermore, the presented fault-aware binning strategy allows accelerators to be binned, with the goal of yield loss reduction, according to the fault-rate and target end-user applications.

\ifCLASSOPTIONcaptionsoff
  \newpage
\fi

\begin{IEEEbiography}[{\includegraphics[width=1in,height=1.25in,clip,keepaspectratio]{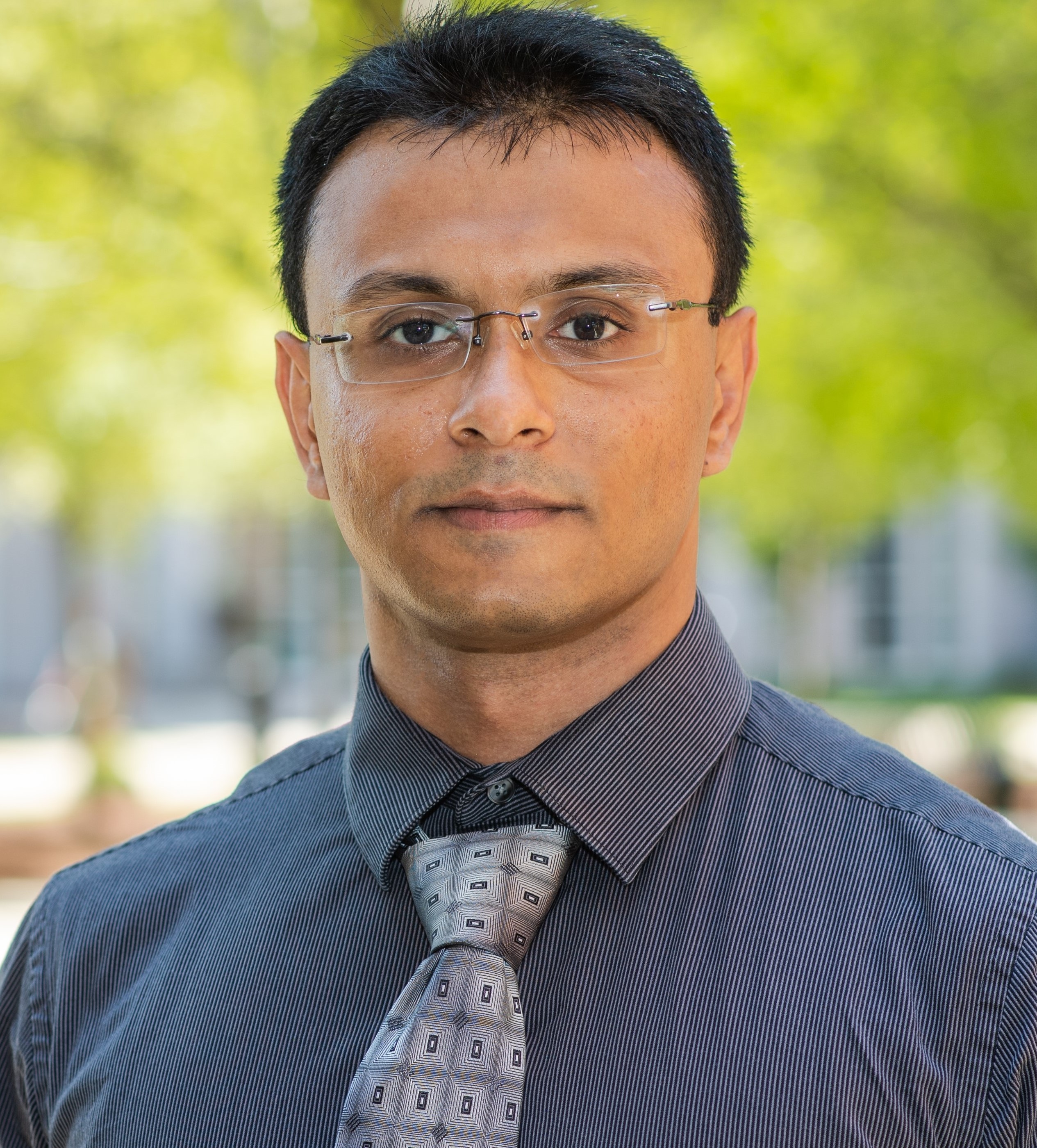}}]{Mehdi Sadi} (S'12-M'17) is currently an Assistant Professor at the Department of Electrical and Computer Engineering (ECE) at Auburn University, Auburn, AL.  Dr. Sadi  earned his PhD in ECE from  University of Florida, Gainesville, USA in 2017, MS from University of California at Riverside, USA in 2011 and BS from Bangladesh University of Engineering and Technology in 2010.   Prior to joining Auburn University, he was a Senior R\&D SoC Design Engineer in the Xeon Design team at Intel Corporation in Oregon. Dr. Sadi`s research focus is on developing algorithms and Computer-Aided-Design (CAD) techniques for implementation, design, test \& reliability of AI, and brain-inspired computing hardware. His research also spans into developing Machine Learning/AI enabled System-on-Chip (SoC) design flows, and Design-for-Reliability for safety-critical AI hardware systems. He has published more than 20 peer-reviewed research papers. He was the recipient of Semiconductor Research Corporation best in session award and Intel Xeon Design Group recognition awards.
\end{IEEEbiography}


\begin{IEEEbiography} [{\includegraphics[width=1in,height=1.25in,clip,keepaspectratio]{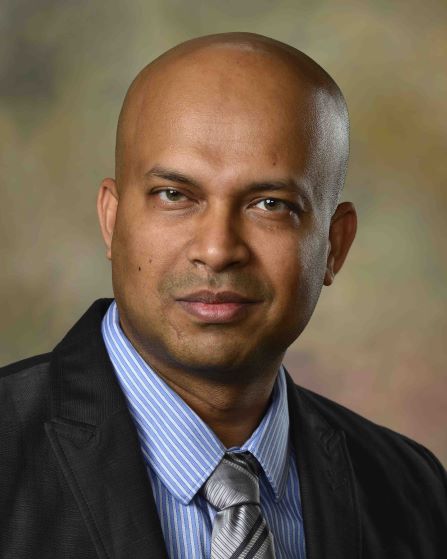}}] {Ujjwal Guin (S'10--M'16)} received his PhD degree from the Electrical and Computer Engineering Department, University of Connecticut, in 2016. He is currently an Assistant Professor in the Electrical and Computer Engineering Department of Auburn University, Auburn, AL, USA. He received his BE degree from the Department of Electronics and Telecommunication Engineering, Bengal Engineering and Science University, Howrah, India, in 2004, and his MS degree from the Department of Electrical and Computer Engineering, Temple University, Philadelphia, PA, USA, in 2010. Dr. Guin has developed several on-chip structures and techniques to improve the security, trustworthiness, and reliability of integrated circuits. His current research interests include Hardware Security \& Trust, Blockchain, Supply Chain Security, Cybersecurity, and VLSI Design \& Test. He is a co-author of the book \textit{Counterfeit Integrated Circuits: Detection and Avoidance}. He has authored several journal articles and refereed conference papers. He was actively involved in developing a web-based tool, Counterfeit Defect Coverage Tool (CDC Tool), \textit{http://www.sae.org/standardsdev/cdctool/}, to evaluate the effectiveness of different test methods used for counterfeit IC detection. He is an active participant in the SAE International G-19A Test Laboratory Standards Development Committee and G-32 Cyber-Physical Systems Security Committee. He is a member of the IEEE and ACM.

\end{IEEEbiography}


{\scriptsize

\begin{thebibliography}{99}
\bibitem{AA1} V. Sze, Y. Chen, T. Yang and J. S. Emer, ``Efficient Processing of Deep Neural Networks: A Tutorial and Survey," in Proceedings of the IEEE, vol. 105, no. 12, pp. 2295-2329, Dec. 2017
\bibitem{AA2} Y. Chen, T. Yang, J. Emer and V. Sze, ``Eyeriss v2: A Flexible Accelerator for Emerging Deep Neural Networks on Mobile Devices," in IEEE Journal on Emerging and Selected Topics in Circuits and Systems, 2019
\bibitem{AA3} J. Lee et. al., ``LNPU: A 25.3TFLOPS/W Sparse Deep-Neural-Network Learning Processor with Fine-Grained Mixed Precision of FP8-FP16,"  IEEE International Solid- State Circuits Conference - (ISSCC), 2019
\bibitem{AA4} N. Jouppi et. al., ``A domain-specific architecture for deep neural networks,’’ Commun. ACM 61, vol 9, pp. 50–59,  2018
\bibitem{AA5} AWS Inferentia Chip: \url{https://aws.amazon.com/}
\bibitem{AA6} S. Markidis, et. al., ``NVIDIA Tensor Core Programmability, Performance \& Precision," IEEE International Parallel and Distributed Processing Symposium Workshops (IPDPSW), 2018
\bibitem{AA7} Gorq Tensor Processor: \url{https://groq.com/technology/}


\bibitem{M1} G. Batra et. al., ``Artificial-intelligence hardware: New opportunities for semiconductor companies,” McKinsey \& Company, January 2019
\bibitem{gpu} Liam Tung, “GPU Killer: Google reveals just how powerful its TPU2 chip really is,” ZDNet, December 14, 2017
\bibitem{cb}  S. Moore, ``Cerebras’s Giant Chip Will Smash Deep Learning’s Speed Barrier,” IEEE Spectrum January, 2020


\bibitem{Time} A. Krizhevsky, I. Sutskever, and G. E. Hinton, `` ImageNet classification with deep convolutional neural networks," in NIPS, 2012
\bibitem{res} K. He et. al, ``Deep Residual Learning for Image Recognition," Computer Vision and Pattern Recognition (CVPR), 2016
\bibitem{vgg} K. Simonyan and A. Zisserman,``Very Deep Convolutional Networks for Large-Scale Image Recognition," in ICLR 2015






\bibitem{img} J. Deng, W. Dong, R. Socher, L. Li, Kai Li and Li Fei-Fei, ``ImageNet: A large-scale hierarchical image database," IEEE CVPR, 2009

\bibitem{Y2} A. J. Strojwas, K. Doong and D. Ciplickas, ``Yield and Reliability Challenges at 7nm and Below," Electron Devices Technology and Manufacturing Conference (EDTM), 2019
\bibitem{Y4} S. Kobayashi, et. al., ``Yield-centric layout optimization with precise quantification of lithographic yield loss," Proc. SPIE, Photomask and Next-Generation Lithography Mask Technology, 2008
\bibitem{Y1} M. Nero, C. Shan, L. Wang and N. Sumikawa, ``Concept Recognition in Production Yield Data Analytics,"  International Test Conference, 2018
\bibitem{Y3} G. Moore et al.,, ``Accelerating 14nm device learning and yield ramp using parallel test structures as part of a new inline parametric test strategy,"  ICMTS, 2015

\bibitem{CA1} P. Maxwell, F. Hapke and H. Tang, ``Cell-aware diagnosis: Defective inmates exposed in their cells,"   European Test Symposium (ETS), 2016
\bibitem{CA2} Z. Gao et al., ``Application of Cell-Aware Test on an Advanced 3nm CMOS Technology Library,"  International Test Conference (ITC), 2019


\bibitem{int} B. Jorgenson, ``Intel’s 2020 Forecast is Grim,” in EE Times, April, 2020

\bibitem{chiplet1} S. Moore, ``3 Ways Chiplets Are Remaking Processors,” IEEE Spectrum, April, 2020
\bibitem{chiplet2} P. Gupta and S. Iyer “Goodbye, Motherboard. Hello, Silicon-Interconnect Fabric,” IEEE Spectrum, October 2019





\bibitem{P1} S. Han, J. Pool, J. Tran, and W.  Dally, ``Learning both weights and connections for efficient neural networks,’’ in International Conference on Neural Information Processing Systems (NIPS’15), 2015
\bibitem{P2} N. Lee, T. Ajanthan and P. Torr, ``SNIP: Single-shot network pruning based on connection sensitivity,’’ in proceedings of International Conference on Learning Representations (ICLR) 2019
\bibitem{P3} S. Han, H. Mao, W.  Dally, ``Deep Compression: Compressing Deep Neural Networks with Pruning, Trained Quantization and Huffman Coding", in ICLR 2016
\bibitem{D1} N. Srivastava et. al., ``Dropout: A Simple Way to Prevent Neural Networks from Overfitting" Journal of Machine Learning Research, 2014














\bibitem{A1} S. Venkataramani, A. Ranjan, K. Roy and A. Raghunathan, ``AxNN: Energy-efficient neuromorphic systems using approximate computing,’’ 2014 IEEE/ACM International Symposium on Low Power Electronics and Design (ISLPED), pp. 27-32, 2014.
\bibitem{A2} Q. Zhang et. al., ``ApproxANN: An approximate computing framework for artificial neural network,’’ 2015 Design, Automation \& Test in Europe Conference \& Exhibition (DATE), pp. 701-706, 2015.
\bibitem{A3} V. Mrazek et. al,, ``Design of power-efficient approximate multipliers for approximate artificial neural networks,’’ in International Conference on Computer-Aided Design (ICCAD ’16), Article 81, 1–7, 2016
\bibitem{A4} M. S. Ansari et. al, ``Improving the Accuracy and Hardware Efficiency of Neural Networks Using Approximate Multipliers,"  IEEE Transactions on Very Large Scale Integration (VLSI) Systems, Feb. 2020.




\bibitem{T3} A. Gebregiorgis and M. B. Tahoori, ``Testing of Neuromorphic Circuits: Structural vs Functional," International Test Conference (ITC), 2019
\bibitem{NY} J. Zhang, K. Basu and S. Garg, ``Fault-Tolerant Systolic Array Based Accelerators for Deep Neural Network Execution," in IEEE Design \& Test, vol. 36, no. 5, pp. 44-53, Oct. 2019.
\textcolor{black}{\bibitem{TE} J. Zhang, K. Rangineni, Z .Ghodsi, and S Garg, ``Thundervolt: enabling aggressive voltage underscaling and timing error resilience for energy efficient deep learning accelerators," in Proceedings of the 55th Annual Design Automation Conference, 2018.}
\textcolor{black}{\bibitem{ME} S. Kim et. al., ``Energy-Efficient Neural Network Acceleration in the Presence of Bit-Level Memory Errors," in IEEE Transactions on Circuits and Systems I: Regular Papers, vol. 65, Dec. 2018.}





\bibitem{sys_f} J. Kim and S. M. Reddy, ``On the design of fault-tolerant two-dimensional systolic arrays for yield enhancement," in IEEE Transactions on Computers, vol. 38, no. 4, pp. 515-525, April 1989.


\bibitem{sim_sys} R.Das and T. Krishna ``DNN Accelerator Architecture – SIMD or Systolic?, " in Computer Architecture Today, ACM SIGARCH, 2018



\bibitem{QQ} J. Wu et. al.,``Quantized Convolutional Neural Networks for Mobile Devices," Computer Vision and Pattern Recognition (CVPR), 2016
\bibitem{int8} S. Migacz, ``8-bit Inference with TensorRT," NVIDIA, 2017
\textcolor{black}{\bibitem{bfloat_1} P. Micikevicius et. al., ``Mixed Precision Training," in ICLR 2018}
\textcolor{black}{\bibitem{bfloat_2} S. Wang and P. Kanwar, ``BFloat16: The secret to high performance on Cloud TPUs," in Google Cloud Blog, 2019}


\bibitem{A5} V. Mrazek et. al.,, ``EvoApprox8b:  Library of Approximate Adders and Multipliers for Circuit Design and Benchmarking of Approximation Methods,’’ in DATE, pp. 258-261, 2017.


\bibitem{core} R. Ray Ramadorai, et. al., ``Method and apparatus for disabling and swapping cores in a multi-core microprocessor,", Intel Corporation, US Patent: US20070255985A1

\bibitem{CNN} online: \url{http://cs231n.stanford.edu/slides/2017/cs231n\_2017\_lecture9.pdf}

\bibitem{DW} Designware IP, online: \url{https://www.synopsys.com/designware-ip.html}
\bibitem{syn} Synopsys: \url{https://www.synopsys.com/}
\bibitem{oc} Opencores: \url{https://opencores.org/}
\bibitem{mat} MATLAB: \url{https://www.mathworks.com/products/matlab.html}
\bibitem{pyt} Pytorch: \url{https://pytorch.org/}


\end{thebibliography}
}
\end{document}